\pdfoutput=0
\documentclass[11pt]{article}
\usepackage{axodraw}
\usepackage{epsfig}
\usepackage{amsfonts}
\usepackage{amsmath}
\usepackage{bm,bbm}
\usepackage{cite}
\allowdisplaybreaks[1]

\input pix.sty

\newcommand{\la}[1]{\label{#1}}
\newcommand{\be}{\begin{equation}}
\newcommand{\ee}{\end{equation}}
\newcommand{\ba}{\begin{eqnarray}}
\newcommand{\ea}{\end{eqnarray}}
\newcommand{\rmi}[1]{{\mbox{\scriptsize #1}}}
\newcommand{\fig}{Fig.~}
\newcommand{\figs}{Figs.~}
\newcommand{\eq}{Eq.~}
\newcommand{\eqs}{Eqs.~}
\newcommand{\se}{Sec.~}
\newcommand{\ses}{Secs.~}
\newcommand{\nr}[1]{(\ref{#1})}

\newcommand{\nn}{\nonumber \\}

\renewcommand{\vec}[1]{{\bf #1}}

\newcommand{\tfr}[2]{{\textstyle \frac{#1}{#2}\,}}

\renewcommand{\eq}{eq.~}
\renewcommand{\eqs}{eqs.~}
\renewcommand{\se}{sec.~}
\renewcommand{\ses}{secs.~}
\renewcommand{\fig}{fig.~}
\renewcommand{\figs}{figs.~}

\newcommand{\alphas}{\alpha_{\rm s}}

\newcommand{\F}{\rmii{$F$}}

\newcommand{\T}{\rmii{$T$}}

\newcommand{\Nf}{N_{\rm f}}
\newcommand{\Nc}{N_{\rm c}}

\newcommand{\Tc}{T_{\rm c}}

\newcommand{\rmO}{{\mathcal{O}}}
\newcommand{\bmu}{\bar\mu}

\def\lsi{\raise0.3ex\hbox{$<$\kern-0.75em\raise-1.1ex\hbox{$\sim$}}}
\def\gsi{\raise0.3ex\hbox{$>$\kern-0.75em\raise-1.1ex\hbox{$\sim$}}}
\newcommand{\lsim}{\mathop{\lsi}}

\newcommand{\nF}{f_\rmii{F}} 
\newcommand{\nB}{f_\rmii{B}} 
\newcommand{\rmii}[1]{{\mbox{\tiny\rm{#1}}}}

\newcommand{\re}{\mathop{\mbox{Re}}}

\newcommand{\Tint}[1]{{\hbox{$\sum$}\!\!\!\!\!\!\!\int\,}_{\!\!\!\!\raise-0.9ex\hbox{$\scriptstyle{#1}$}}}
\newcommand{\Tinti}[1]{{{\Sigma}\!\!\!\!\raise0.3ex\hbox{$\int$}_\rmii{${#1}$}}}

\newcommand{\bi}{\begin{itemize}}
\newcommand{\ei}{\end{itemize}}
\newcommand{\hide}[1]{ }
\newcommand{\blind}[1]{\fbox{$ ? $}} 

\newcommand{\deltabar}{\raise-0.02em\hbox{$\bar{}$}\hspace*{-0.8mm}{\delta}}
\newcommand{\fullscat}[1]{\mbox{aver}^{ }_\rmi{$#1$}}
\newcommand{\scat}[1]{\mbox{scat}^{ }_\rmi{$#1$}}
\newcommand{\s}[1]{s_{#1}}
\renewcommand{\P}{\mathcal{P}}
\newcommand{\K}{\mathcal{K}}
\newcommand{\aS}{\varphi} 
\newcommand{\bS}{\widetilde\varphi} 
\newcommand{\aZ}{\rmii{$Z$}}
\newcommand{\aW}{\rmii{$W$}}
\newcommand{\mS}{m_\aS}
\newcommand{\inel}{\rmi{inel}}
\newcommand{\elas}{\rmi{elas}}
\newcommand{\vrel}{v^{ }_\rmi{rel}}
\newcommand{\real}{real}
\newcommand{\virt}{virt}
\def\TAsc(#1,#2)(#3,#4,#5)%
{\SetWidth{2.0}\CArc(#1,#2)(#3,#4,#5)\SetWidth{1.0}}
\def\Lwidth{3}

\def\TAgl(#1,#2)(#3,#4,#5){\SetWidth{2.0}\PhotonArc(#1,#2)(#3,#4,#5){\Lwidth}%
{6.283 #3 mul 360 div #4 #5 sub #4 #5 sub mul sqrt mul Tdensity mul}%
\SetWidth{1.0}}
\def\TLgl(#1,#2)(#3,#4){\SetWidth{2.0}\Photon(#1,#2)(#3,#4){\Lwidth}
{#1 #3 sub #1 #3 sub mul #2 #4 sub #2 #4 sub mul add sqrt Tdensity mul}%
\SetWidth{1.0}}

\renewcommand{\picc}[1]{\;\parbox[c]{60pt}{\begin{picture}(60,30)(0,-10)
\SetWidth{1.0}\SetScale{1.3} #1 \end{picture}}\; }
\def\Lwidth{1.3}

\newcommand{\picu}[1]{\;\parbox[c]{70pt}{\begin{picture}(70,30)(-15,-5)
\SetWidth{1.0}\SetScale{0.65} #1 \end{picture}}\; }

\def\Ampl{\picu{%
 \Line(40,0)(20,15)%
 \Lsc(40,30)(20,15)%
 \COval(20,15)(2,2)(0){Black}{Black}%
 \Line(18,16.5)(0,16.5)%
 \Line(18,13.5)(0,13.5)
}}
\def\AmplD{\picu{%
 \Line(40,0)(20,15)%
 \Lsc(20,15)(30,22.5)%
 \Laqu(40,30)(30,22.5)%
 \Lqu(40,15)(30,22.5)%
 \COval(20,15)(2,2)(0){Black}{Black}%
 \Line(18,16.5)(0,16.5)%
 \Line(18,13.5)(0,13.5)
}}
\def\AmplE{\picu{%
 \Line(40,0)(20,15)%
 \Lsc(20,15)(30,22.5)%
 \Photon(30,22.5)(40,30){1.5}{2.5}%
 \Photon(30,22.5)(40,15){-1.5}{2.5}%
 \COval(20,15)(2,2)(0){Black}{Black}%
 \Line(18,16.5)(0,16.5)%
 \Line(18,13.5)(0,13.5)
}}
\def\AmplF{\picu{%
 \Line(40,0)(20,15)%
 \Lsc(20,15)(30,22.5)%
 \Gluon(30,22.5)(40,30){1.8}{2}%
 \Gluon(30,22.5)(40,15){-1.8}{2}%
 \COval(20,15)(2,2)(0){Black}{Black}%
 \Line(18,16.5)(0,16.5)%
 \Line(18,13.5)(0,13.5)
}}
\def\AmplG{\picu{%
 \Line(40,0)(20,15)%
 \Lsc(20,15)(30,22.5)%
 \Lsc(30,22.5)(40,30)%
 \Lsc(30,22.5)(40,15)%
 \COval(20,15)(2,2)(0){Black}{Black}%
 \Line(18,16.5)(0,16.5)%
 \Line(18,13.5)(0,13.5)
}}
\def\AmplH{\picu{%
 \Line(40,0)(20,15)%
 \Lsc(20,15)(40,30)%
 \Lsc(20,15)(40,15)%
 \COval(20,15)(2,2)(0){Black}{Black}%
 \Line(18,16.5)(0,16.5)%
 \Line(18,13.5)(0,13.5)
}}
\def\AmplI{\picu{%
 \Lsc(20,15)(40,30)%
 \Lsc(30,7.5)(40,15)%
 \Line(20,15)(40,0)%
 \COval(30,7.5)(2,2)(0){Black}{Black}%
 \COval(20,15)(2,2)(0){Black}{Black}%
 \Line(18,16.5)(0,16.5)%
 \Line(18,13.5)(0,13.5)
}}
\def\AmplJ{\picu{%
 \Lsc(20,15)(40,15)%
 \Line(30,7.5)(40,0)%
 \Line(20,15)(30,7.5)%
 \Lsc(30,7.5)(32.67,13.5)%
 \Lsc(34,16.5)(40,30)%
 \COval(30,7.5)(2,2)(0){Black}{Black}%
 \COval(20,15)(2,2)(0){Black}{Black}%
 \Line(18,16.5)(0,16.5)%
 \Line(18,13.5)(0,13.5)
}}
\def\AmplK{\picu{%
 \Line(40,0)(20,15)%
 \Line(20,15)(40,30)%
 \Line(20,15)(40,15)%
 \COval(20,15)(2,2)(0){Black}{Black}%
 \Line(18,16.5)(0,16.5)%
 \Line(18,13.5)(0,13.5)
}}
\def\AmplL{\picu{%
 \Line(40,0)(20,15)%
 \Lsc(20,15)(30,22.5)%
 \Line(30,22.5)(40,30)%
 \Line(30,22.5)(40,15)%
 \COval(30,22.5)(2,2)(0){Black}{Black}%
 \COval(20,15)(2,2)(0){Black}{Black}%
 \Line(18,16.5)(0,16.5)%
 \Line(18,13.5)(0,13.5)
}}
\def\AmplM{\picu{%
 \Line(20,15)(40,30)%
 \Line(30,7.5)(40,15)%
 \Lsc(20,15)(30,7.5)%
 \Line(30,7.5)(40,0)%
 \COval(30,7.5)(2,2)(0){Black}{Black}%
 \COval(20,15)(2,2)(0){Black}{Black}%
 \Line(18,16.5)(0,16.5)%
 \Line(18,13.5)(0,13.5)
}}
\def\AmplN{\picu{%
 \Line(20,15)(40,15)%
 \Line(30,22.5)(40,30)%
 \Lsc(20,15)(30,22.5)%
 \Line(30,22.5)(32.67,16.5)%
 \Line(34,13.5)(40,0)%
 \COval(30,22.5)(2,2)(0){Black}{Black}%
 \COval(20,15)(2,2)(0){Black}{Black}%
 \Line(18,16.5)(0,16.5)%
 \Line(18,13.5)(0,13.5)
}}
\def\AmplO{\picu{%
 \Line(40,0)(20,15)%
 \Lsc(40,30)(20,15)%
 \Line(18,16.5)(0,16.5)%
 \Line(18,13.5)(0,13.5)%
 \COval(20,15)(5,5)(0){Black}{Gray}
}}
\def\AmplP{\picu{%
 \Line(40,0)(20,15)%
 \Lsc(40,30)(20,15)%
 \Line(18,16.5)(0,16.5)%
 \Line(18,13.5)(0,13.5)%
 \COval(20,15)(2,2)(0){Black}{Black}%
 \COval(30,22.5)(4,4)(0){Black}{Gray}
}}
\def\AmplQ{\picu{%
 \Line(40,0)(20,15)%
 \Lsc(40,30)(20,15)%
 \Line(18,16.5)(0,16.5)%
 \Line(18,13.5)(0,13.5)%
 \COval(20,15)(2,2)(0){Black}{Black}%
 \COval(30,7.5)(4,4)(0){Black}{Gray}
}}
\def\VertexA{\picu{%
 \Line(40,0)(20,15)%
 \Lsc(40,30)(20,15)%
 \Lsc(30,22.5)(30,7.5)%
 \COval(20,15)(2,2)(0){Black}{Black}%
 \COval(30,7.5)(2,2)(0){Black}{Black}%
 \Line(18,16.5)(0,16.5)%
 \Line(18,13.5)(0,13.5)
}}
\def\VertexB{\picu{%
 \Lsc(20,15)(30,7.5)%
 \Line(30,7.5)(40,0)%
 \Lsc(30,22.5)(40,30)%
 \Line(20,15)(30,22.5)%
 \Line(30,22.5)(30,7.5)%
 \COval(20,15)(2,2)(0){Black}{Black}%
 \COval(30,7.5)(2,2)(0){Black}{Black}%
 \COval(30,22.5)(2,2)(0){Black}{Black}%
 \Line(18,16.5)(0,16.5)%
 \Line(18,13.5)(0,13.5)
}}
\def\VertexC{\picu{%
 \Line(40,0)(20,15)%
 \Lsc(40,30)(20,15)%
 \Asc(23,5)(10,0,110)%
 \COval(20,15)(2,2)(0){Black}{Black}%
 \COval(32,6)(2,2)(0){Black}{Black}%
 \Line(18,16.5)(0,16.5)%
 \Line(18,13.5)(0,13.5)
}}
\def\VertexD{\picu{%
 \Line(40,0)(20,15)%
 \Lsc(40,30)(20,15)%
 \Asc(25,24.5)(10,250,360)%
 \COval(20,15)(2,2)(0){Black}{Black}%
 \Line(18,16.5)(0,16.5)%
 \Line(18,13.5)(0,13.5)
}}
\def\VertexE{\picu{%
 \Line(40,0)(20,15)%
 \Lsc(40,30)(32,24)%
 \Line(32,24)(20,15)%
 \CArc(25,24.5)(9.5,250,360)%
 \COval(20,15)(2,2)(0){Black}{Black}%
 \COval(33,24.75)(2,2)(0){Black}{Black}%
 \Line(18,16.5)(0,16.5)%
 \Line(18,13.5)(0,13.5)
}}
\def\VertexF{\picu{%
 \Line(49,0)(34,15)%
 \Lsc(49,30)(34,15)%
 \Asc(27,13)(9,10,170)%
 \CArc(27,17)(9,190,350)%
 \COval(19.5,15)(2,2)(0){Black}{Black}%
 \COval(35,15)(2,2)(0){Black}{Black}%
 \Line(18,16.5)(0,16.5)%
 \Line(18,13.5)(0,13.5)
}}
\def\selfEA{\picu{%
 \Lsc(0,15)(20,15)%
 \Lsc(40,15)(20,15)%
 \COval(20,15)(5,5)(0){Black}{Gray}
}}
\def\selfEB{\picu{%
 \Line(0,15)(20,15)%
 \Line(40,15)(20,15)%
 \COval(20,15)(5,5)(0){Black}{Gray}
}}
\def\scalarA{\picu{%
 \Lsc(0,15)(12,15)%
 \Lsc(40,15)(28,15)%
 \Asc(20,15)(8,0,180)%
 \Asc(20,15)(8,180,360)%
}}
\def\scalarB{\picu{%
 \Lsc(0,15)(12,15)%
 \Lsc(40,15)(28,15)%
 \Asc(20,15)(8,0,180)%
 \Agl(20,15)(8,180,360)%
}}
\def\scalarC{\picu{%
 \Lsc(0,15)(12,15)%
 \Lsc(40,15)(28,15)%
 \Agl(20,15)(8,0,180)%
 \Agl(20,15)(8,180,360)%
}}
\def\scalarG{\picu{%
 \Lsc(0,15)(12,15)%
 \Lsc(40,15)(28,15)%
 \Aagh(20,15)(8,0,180)%
 \Aagh(20,15)(8,180,360)%
}}
\def\scalarD{\picu{%
 \Lsc(0,15)(12,15)%
 \Lsc(40,15)(28,15)%
 \CArc(20,15)(8,0,180)%
 \CArc(20,15)(8,180,360)%
 \COval(12,15)(2,2)(0){Black}{Black}%
 \COval(28,15)(2,2)(0){Black}{Black}%
}}
\def\scalarE{\picu{%
 \Lsc(0,15)(12,15)%
 \Lsc(40,15)(28,15)%
 \Aaqu(20,15)(8,0,180)%
 \Aaqu(20,15)(8,180,360)%
}}
\def\scalarF{\picu{%
 \Lsc(0,15)(12,15)%
 \Lsc(40,15)(28,15)%
 \GlueArc(20,15)(7,0,180){2.2}{3}%
 \GlueArc(20,15)(7,180,360){2.2}{3}%
}}
\def\scalarH{\picu{%
 \Lsc(0,15)(20,13)%
 \Lsc(40,15)(20,13)%
 \Asc(20,21)(8,-90,270)%
}}
\def\scalarI{\picu{%
 \Lsc(0,15)(20,13)%
 \Lsc(40,15)(20,13)%
 \Agl(20,21)(8,-90,270)%
}}
\def\scalarJ{\picu{%
 \Lsc(0,15)(20,13)%
 \Lsc(40,15)(20,13)%
 \CArc(20,21)(8,-90,270)%
 \COval(20,13)(2,2)(0){Black}{Black}%
}}
\def\singletA{\picu{%
 \Line(0,15)(12,15)%
 \Line(40,15)(28,15)%
 \Asc(20,15)(8,0,180)%
 \CArc(20,15)(8,180,360)%
 \COval(12,15)(2,2)(0){Black}{Black}%
 \COval(28,15)(2,2)(0){Black}{Black}%
}}
\def\singletB{\picu{%
 \Line(0,15)(20,13)%
 \Line(40,15)(20,13)%
 \Asc(20,21)(8,-90,270)%
 \COval(20,13)(2,2)(0){Black}{Black}%
}}
\def\singletC{\picu{%
 \Line(0,15)(20,13)%
 \Line(40,15)(20,13)%
 \CArc(20,21)(8,-90,270)%
 \COval(20,13)(2,2)(0){Black}{Black}%
}}

\makeatletter \@addtoreset{equation}{section} \makeatother
\renewcommand{\theequation}{\arabic{section}.\arabic{equation}}
\makeatletter
\renewcommand\section{\@startsection {section}{1}{\z@}%
                                   {-5.5ex \@plus -1ex \@minus -.2ex}
                                   {2.3ex \@plus.2ex}%
                                   {\normalfont\large\bfseries}}
\renewcommand\subsection{\@startsection{subsection}{2}{\z@}%
                                     {-3.25ex\@plus -1ex \@minus -.2ex}%
                                     {1.5ex \@plus .2ex}%
                                     {\normalfont\normalsize\bfseries}}
\renewcommand\thesection {\@arabic\c@section}
\renewcommand\thesubsection   {\thesection.\@arabic\c@subsection}
\renewcommand{\@seccntformat}[1]{%
\csname the#1\endcsname.\hspace{1.0em}}
\makeatother

\begin{document}

\flushbottom

\begin{titlepage}

\begin{flushright}
January 2023
\end{flushright}
\begin{centering}
\vfill

{\Large{\bf
  Resonant $s$-channel dark matter annihilation at NLO 
}} 

\vspace{0.8cm}

M.~Laine 

\vspace{0.8cm}

{\em
AEC, 
Institute for Theoretical Physics, 
University of Bern, \\ 
Sidlerstrasse 5, CH-3012 Bern, Switzerland \\}

\vspace*{0.8cm}

\mbox{\bf Abstract}
 
\end{centering}

\vspace*{0.3cm}
 
\noindent
Studies of dark matter annihilation through an $s$-channel resonance
are often based on recipes such as a narrow width approximation or
real intermediate state subtraction. We review a recipe-free formalism
that can be implemented at the NLO level in the full theory, and
ensures the cancellation of mass singularities.  Its basic ingredients
can be formulated in the relativistic regime, but we show that the
procedure simplifies if we go to the non-relativistic one and assume
the presence of kinetic equilibrium.  The latter case is illustrated
for scalar singlet dark matter with $\mS^{ }\simeq 60$~GeV, freezing
out at $T \simeq (1-3)$~GeV, re-confirming the viability of this
scenario with couplings tiny enough to evade experimental constraints.

\vfill


\end{titlepage}

\tableofcontents

%
\section{Introduction}
\la{se:intro}

As traditional dark matter scenarios are put under pressure
by collider searches and direct and indirect observational constraints, 
refined frameworks may become interesting. 
One possibility originates through ``resonant'' 
effects~\cite{old1,old2}. Notably, the resonance
could originate via an attractive 
$t$-channel exchange of light
force carriers, leading to 
a large enhancement in the spectral 
density of low-energy scattering states
and possibly even to bound states between dark sector particles; 
or it could be an elementary excitation created
in the $s$-channel, with a mass larger than twice the 
dark matter mass. Both of these could lead to an
efficient depletion of dark matter particles
in the early universe, whereby
the correct abundance could be reached via the 
freeze-out mechanism, despite tiny couplings to the Standard Model. 

To make the point concrete, 
we recall that among 
the simplest dark matter models is Standard Model
extended by 
a singlet scalar field~\cite{singlet1,singlet2,singlet3,singlet4}.
As this setup has been studied in increasing detail, 
a curious corner of parameter space has been identified, 
with a small singlet mass 
$m^{ }_{\aS}\sim 60$~GeV $< m^{ }_h/2$ and very weak couplings,
which is not phenomenologically excluded, 
despite the weak-scale mass~(cf.,\ e.g., ref.~\cite{aa}
and references therein). 

Curious corners of parameter space 
sometimes invoke non-trivial physics. 
Indeed, for the said example, 
a substantial argument has emerged between 
two groups~\cite{tb,kk,drake,kk2}, concerning the role that
kinetic equilibrium plays in this scenario. This has led, amongst
others, to recipes for treating
kinetic non-equilibrium in a numerically more 
manageable manner~\cite{tb,kk,drake,abe,kin,kin2,kk2}. 

However, apart from kinetic non-equilibrium, 
there could be other reasons for uncertainties in existing
computations. One issue is that, 
as exemplified by ref.~\cite{singlet4}, 
the annihilation cross section of scalar singlet particles to Standard Model
particles has been estimated by giving the Higgs propagator a finite width, 
namely its well-known vacuum decay rate. 
Though a reasonable approximation numerically, 
this is conceptually unsatisfactory,
given that the Bose enhancement
or Pauli blocking factors 
of the final-state particles in a thermal environment
are omitted.\footnote{%
 In vacuum, it has been proposed that this approximation can be 
 systematized into an effective field theory~\cite{eft}, 
 however its application to a general thermal environment is unclear, 
 since the center-of-mass frame of the 2-particle final state 
 differs from the plasma rest frame. 
 } 
Indeed the Higgs width is known to be modified 
by thermal corrections~\cite{gw}.
Another problem is that dark matter freeze-out takes place 
in the temperature regime $T \sim (1-3)$~GeV, where poorly 
understood QCD effects could be substantial. 

The purpose of the present paper is to address the latter uncertainties
from a somewhat more general perspective, 
while not (yet) tackling the issue of kinetic non-equilibrium. 
The basic point is that in an unresummed order-by-order 
computation, an $s$-channel resonance is to be treated as 
an on-shell particle. The dominant 
dark matter annihilation channel is then the $2\to 1$
``inverse decay'' of two singlet scalars into an on-shell Higgs. 
The Higgs decays, 
represented by the width, are next-to-leading order 
(NLO) reactions, such as $2\to 2$. 
However, at the same order, virtual corrections to the $2\to 1$
process should be included, and are in fact crucial, 
as they cancel mass singularities 
according to the KLN theorem~\cite{kln1,kln2}.
By formulating the theoretical side by a consistent NLO 
treatment including these effects, we may also hope to 
incorporate thermal QCD effects in a somewhat reasonable manner. 

Our presentation is organized as follows. 
Most dark matter computations adopt Boltzmann equations
as their starting point. 
In \se\ref{se:boltzmann}, we recall why 
text-book Boltzmann equations
provide an incomplete treatment of nature 
when proceeding towards the NLO level, 
and one way to rectify them 
by a quantum-field theoretic computation. 
Moreover the simplifications met in the non-relativistic regime
and in the presence of kinetic equilibrium are spelled out. 
In \se\ref{se:MstarM}, the ingredients needed for implementing 
the NLO treatment are summarized for the scalar singlet model, 
with details of matrix elements squared relegated to appendix~A, 
and leading-order phase-space integrals to appendix~B. 
Our numerical results are presented in \se\ref{se:numerics}.
We turn to a summary in \se\ref{se:concl}, adding
at the same time a proposal on how the issue of
kinetic non-equilibrium might be attacked beyond
the Boltzmann level. 

\section{Boltzmann equation and how to go beyond it}
\la{se:boltzmann}

\subsection{Basic setup and one of its deficiencies}
\la{se:basics}

Denoting by $\K \equiv (\omega,\vec{k})$ the four-momentum of a dark 
matter particle, which is here assumed to be a boson of mass $m_\aS^{ }$, 
so that $\omega \equiv \sqrt{k^2 + m_\aS^2}$ 
where $k \equiv |\vec{k}|$; 
and by $f^{ }_{\aS}$ its phase space density, the Boltzmann 
equation governing the dark matter 
evolution has in local Minkowskian coordinates in the plasma
rest frame the form 
\be 
 \mathcal{K}^\alpha \partial^{ }_\alpha f^{ }_{\aS}
 \; = \; 
 - \sum_{m,n} 
 \fullscat{1+m\to n}(a^{ }_1,...,a^{ }_m;b^{ }_1,...,b^{ }_n) 
 \, 
 c \sum_\rmi{spins} |\mathcal{M}|^2_{{\aS}+m\to n}
 \;, \la{boltzmann}
\ee 
where we have assumed the ``mostly minus'' metric convention; 
the $a^{ }_i$ label particles in the 
initial state of the loss term (in addition to $\aS$); 
the $b^{ }_i$ stand for final-state particles in the loss term; 
and the phase space average has been defined as 
\ba
 && \hspace*{-1.0cm} 
 \fullscat{1+m\to n}(a^{ }_1,...,a^{ }_m;b^{ }_1,...,b^{ }_n) \;\equiv\;
 \frac{1}{2} 
 \int \! {\rm d}\Phi^{ }_{1 + m \to n} 
 \nn & \times & 
 \bigl \{  
 f^{ }_{\aS} f^{ }_{a_1} \cdots f^{ }_{a_m} 
 (1\pm f^{ }_{b_1}) \cdots (1\pm f^{ }_{b_n}) 
 - 
 f^{ }_{b_1} \cdots f^{ }_{b_n} 
 (1 + f^{ }_{\aS})(1\pm f^{ }_{a_1}) \cdots (1\pm f^{ }_{a_m})
 \bigr\}
 \;. \hspace*{5mm} \la{fullscat}
\ea
The phase space integral goes over the momenta of the particle sets
$\{a^{ }_i\}$ and $\{b^{ }_i\}$, 
and the signs $\pm$ apply to bosons and fermions, respectively. 
The factor $c \equiv 1 / ( i^{ }_a! i^{ }_b!)$
in \eq\nr{boltzmann} cancels overcounting when
integrating over the momenta of $i^{ }_a$ or $i^{ }_b$
identical particles in
the initial or final state. 
The sum $\sum_\rmi{spins}$ 
in \eq\nr{boltzmann} goes over polarizations, 
and we have assumed the symmetry
$
 |\mathcal{M}|^2_{{\aS}+m\to n} = 
 |\mathcal{M}|^2_{n\to {\aS}+m}
$
for the matrix elements squared.

In the following, we assume that 
all Standard Model particles are in thermal equilibrium, 
so that their phase space distribution $f$ can be replaced 
by the Bose ($\equiv \nB^{ }$)
or Fermi distribution ($\equiv \nF^{ }$). 
For equilibrated particles, 
$ 1 \pm f^{ }_a = e^{\beta \epsilon^{ }_a} f^{ }_a$, 
where $\beta \equiv 1/T$ and the momenta were written as 
$\P^{ }_a \equiv (\epsilon^{ }_a,\vec{p}^{ }_a)$.
If the $\aS$-particles were also in full equilibrium, 
then energy conservation, 
$
 \omega + \epsilon^{ }_{a_1} + ... + \epsilon^{ }_{a_m}
 = 
 \epsilon^{ }_{b_1} + ... + \epsilon^{ }_{b_n}
$,
would guarantee detailed balance, 
i.e.\ that the right-hand side of \eq\nr{fullscat} vanishes. 

In the physical situation, the $\aS$-particles may fall out of 
chemical and/or kinetic equilibrium. 
Thereby \eq\nr{boltzmann} turns into
an integro-differential equation for $f^{ }_{\varphi}$. 
Even though such equations can be 
solved, by discretizing momentum space (and, if the system is not 
translationally invariant, configuration space as well), 
the solution tends to be numerically expensive. Furthermore, 
such a solution does 
{\em not} represent an exact treatment of nature. 

To appreciate the latter point, we recall that
one deficiency of the Boltzmann equation 
in \eq\nr{boltzmann} is that its building blocks are what
we call {\em real processes}, between on-shell particles
whose phase-space distributions we know or want to determine. 
{\em Virtual corrections}
(closed loops) can only be incorporated in so far as they amount to 
vacuum corrections to 
the matrix elements squared. But virtual corrections involving thermal
effects --- for instance, thermal corrections to masses or couplings, 
or more generally thermal corrections to dispersion relations --- 
are not present. Yet this can be important, for
instance by opening up new channels that would not be allowed 
by vacuum kinematics. 

It is for this reason that for 
a systematic treatment, the Boltzmann equation needs to be replaced
by a quantum field theoretic description, in which both real and
virtual processes, as well as all cancellations between them, are
automatically present. It is not clear, {\em a priori}, how this 
can be achieved in general 
(though specific examples have been worked out, 
see e.g.\ ref.~\cite{mb}). 
However, one transparent possibility is if we can 
define coefficients that amount to {\em equilibration rates}, 
which have an unambiguous physical meaning 
in the linear response regime. 
In the present paper, we show how this 
can be achieved through the definitions
of a somewhat formal construction that we
call the maximal interaction rate, 
and the physically important chemical equilibration rate, 
in \ses\ref{se:interaction}
and \ref{se:chemical}, respectively. 

\subsection{Maximal interaction rate}
\la{se:interaction}

The purpose of the present section is to manipulate the 
Boltzmann equation in \eq\nr{boltzmann} in order to identify
what we term the maximal interaction rate. 
It should be stressed from the outset 
that the result is not inherent to a Boltzmann equation, 
but more general. In other words, 
the assumptions we make for its derivation
are sufficient but not necessary. 
For instance, in ref.~\cite{dbx}, 
the same rate 
and rate equation were obtained from quantum field theory, 
by carrying out an analysis 
to leading order in a weak coupling between a $\varphi$-field
and Standard Model, but to all orders in Standard
Model couplings. 

Let us assume for a moment 
that $f^{ }_{\aS}$ is close to equilibrium, 
apart from around the momentum bin $\vec{k}$, and expand 
to first order in deviations in this bin, {\it viz.}  
\be
 f^{ }_{\aS} = \bar{f}^{ }_\aS + \delta f^{ }_{\aS}
 \;, \quad
 \bar{f}^{ }_\aS(k) \; \equiv \; \nB^{ }(\omega)
 \;, \quad
 \delta f^{ }_{\aS}({k}) \;\equiv\; f^{ }_{\aS}({k}) - \nB^{ }(\omega) 
%
 \;. \la{kin_dev}
\ee
Given that a single momentum bin
can be excluded from the integrations 
over $\P^{ }_{a_i}$ and $\P^{ }_{b_i}$ 
in \eq\nr{fullscat} without 
significantly affecting the outcome, 
$f^{ }_{\aS}$ can be replaced by $\nB^{ }$
if it appears in the sets $\{ a^{ }_i \}$ or $\{ b^{ }_i \}$.
Recalling furthermore 
that the zeroth order term vanishes by detailed balance, 
it follows from \eq\nr{fullscat} that, 
to first order in $\delta f^{ }_{\aS}$,  
\ba
 && \hspace*{-1.0cm} 
 \fullscat{1+m\to n}(a^{ }_1,...,a^{ }_m;b^{ }_1,...,b^{ }_n)
 \nn  & = & 
 \delta f^{ }_{\aS}  \times \frac{1}{2} 
 \int \! {\rm d}\Phi^{ }_{1 + m \to n} 
 \, 
 \Bigl \{  
 f^{ }_{\sigma_{a_1}} \cdots f^{ }_{\sigma_{a_m}} 
 (1 + f^{ }_{\sigma_{b_1}}) \cdots (1 + f^{ }_{\sigma_{b_n}}) 
 \nn & & \hspace*{3.4cm} -  \, 
 f^{ }_{\sigma_{b_1}} \cdots f^{ }_{\sigma_{b_n}} 
 (1 + f^{ }_{\sigma_{a_1}}) \cdots (1 + f^{ }_{\sigma_{a_m}})
 \Bigr\} \, (-1)^\F 
 \; + \; \rmO(\delta f^{2}_{\aS})
 \nn[2mm]
 & \equiv & 
 \bigl[ f^{ }_{\aS}({k}) - \nB^{ }(\omega) \bigr]
 \times
 \scat{1+m\to n}(-a^{ }_1,...,-a^{ }_m;b^{ }_1,...,b^{ }_n)
 \, (-1)^{\F}_{ } 
 \; + \; \rmO(\delta f^{2}_{\aS})
 \;. 
 \la{scat} 
\ea
Here $\sigma^{ }_i = \pm$ 
denotes the statistics of each particle species; 
we have introduced 
$f^{ }_+ \equiv \nB^{ }$, 
$f^{ }_- \equiv - \nF^{ }$; 
$F$ is the number of fermions in 
the initial (or final) state; 
$ (-1)^{\F}_{ } $ is a factor originating from 
the sign difference between $f^{ }_-$ and $\nF^{ }$; 
and 
$ \scat{1+m\to n} $
corresponds to the notation 
introduced in ref.~\cite{phasespace}.  

The rationale for introducing negative signs in front of the particle
labels in the argument of $\scat{1+m\to n}$ in \eq\nr{scat} 
is that $\scat{1+m\to n}$ can be defined as an operator, 
such that negative labels
invert the signs of the corresponding
momenta in the matrix element squared. Thereby all matrix elements squared
can be obtained by crossings from a would-be decay matrix element squared, 
which enjoys maximal symmetries.
Specifically, defining 
\be
 \Theta(\P^{ }_{a_1},...,\P^{ }_{a_m},\P^{ }_{b_1},...,\P^{ }_{b_n})
 \; \equiv \; 
 c \sum_\rmi{spins} |\mathcal{M}|^2_{{\aS}\to m + n}
 \;, \la{Theta_def}
\ee
where all momenta are now in the final state,
the combination originating from 
\eqs\nr{boltzmann} and \nr{scat} amounts to   
\be
 (-1)^\F \,  
 c \sum_\rmi{spins} |\mathcal{M}|^2_{{\aS}+m\to n}
 = 
 \Theta(-\P^{ }_{a_1},...,-\P^{ }_{a_m},\P^{ }_{b_1},...,\P^{ }_{b_n})
 \;.
\ee
Adopting the operator notation, 
we define the real-scattering part of 
an interaction rate as
\be
 \omega\,\Gamma^\rmi{\real}_\rmi{max}(k) \; \equiv \; 
 \sum_{m,n} \scat{1+m\to n} (-a^{ }_1,...,-a^{ }_m;b^{ }_1,...,b^{ }_n)
        \,    \Theta(\P^{ }_{a_1},...,\P^{ }_{a_m},
                     \P^{ }_{b_1},...,\P^{ }_{b_n})
 \;. \la{phasespace}
\ee

Now, if \eq\nr{phasespace} originates from a quantum field theoretic
equilibration rate, 
in the sense described in ref.~\cite{dbx}, where it corresponds
to the imaginary part of a retarded self-energy, 
then the full rate includes also {\em virtual processes}. 
We denote the latter by 
$
  \Gamma^\rmi{\virt}_\rmi{max}
$.
At NLO, the physical, and thereby ultraviolet (UV) 
and infrared (IR) finite rate, is given by
\be
 \Gamma^\rmi{phys}_\rmi{max}(k) \; = \; 
 \Gamma^\rmi{\real}_\rmi{max}(k) \; + \; 
 \Gamma^\rmi{\virt}_\rmi{max}(k)
 \;.  \la{max}
\ee
Going over to an expanding background, with the Hubble rate
denoted by $H \; \equiv \; \dot{a}/a$,
and assuming furthermore that $f^{ }_\aS$ is translationally invariant, 
\eq\nr{boltzmann} then takes the form 
\be
 \bigl( \partial^{ }_t - H k\, \partial^{ }_k \bigr) f^{ }_\aS (k) 
 \; = \; - \Gamma^\rmi{phys}_\rmi{max}(k) \, 
 \bigl[ f^{ }_\aS(k) - \nB^{ }(\omega) \bigr]  + 
 \rmO(\delta f^{2}_\aS)
 \;. \la{boltzmann2}
\ee
This equation can be viewed as describing 
equilibration in the sense of linear response theory. 
On the other hand, 
by setting $f^{ }_{\aS} \to 0$ on the right-hand side, 
it defines the production rate of the $\aS$ particles from 
a plasma. Both interpretations underline that 
the equation has a physical meaning 
beyond Boltzmann equations. 

A cautionary word needs to be added, however. 
Even though \eq\nr{boltzmann2} 
shows that $ \Gamma^\rmi{phys}_\rmi{max} $
drives the system towards the
Bose distribution if the system is already close to it, 
$ \Gamma^\rmi{phys}_\rmi{max} $ should {\em not} be
interpreted as a kinetic equilibration rate. 
Indeed kinetic
equilibration is a notion associated with 
particles whose
phase-space distribution can differ from 
equilibrium by an overall factor
(if the particles are out of chemical equilibrium)
and by a different shape. Kinetic equilibration
involves transfer of momentum, in order to rectify the shape, 
but the overall normalization should 
not be simultaneously changed, as it 
represents the number density. Instead, 
\eq\nr{boltzmann2} represents the maximal rate at which
the $\aS$-particles interact; 
apart from a change of momentum or particle number, 
this includes the very fast processes that
lead to phase decoherence in quantum mechanics.

\subsection{Chemical equilibration rate}
\la{se:chemical}

In the previous section 
we defined a rate from Boltzmann equations which
can arguably
be generalized to have a quantum field theoretic meaning,
as the imaginary part of a retarded self-energy
in the linear response regime. We start
the present section by recalling that in the non-relativistic
limit such rates have a well-defined subpart, which then also has a 
quantum field theoretic meaning. Subsequently, returning to Boltzmann
equations, we show that the momentum average of this subpart has 
an interpretation as the chemical equilibration rate.\footnote{%
 In the non-relativistic limit a chemical equilibration rate can also be 
 defined directly in quantum field theory~\cite{chem}, with the
 connection to Boltzmann equations then following in the course of
 its practical evaluation. 
 }

Let us separate all possible scatterings 
into two classes, 
according to whether the number of $\aS$ particles changes 
in the reaction (``inelastic processes''), or not
(``elastic processes''),  {\it viz.} 
\be
 \Gamma^\rmi{phys}_\rmi{max}(k) 
 \; \equiv \; 
 \Gamma^\rmi{phys}_\inel(k)
 + 
 \Gamma^\rmi{phys}_\elas(k)
 \;. \la{inel_elas}
\ee
The same division can be made 
separately
in 
$
 \Gamma^\rmi{\real}_\rmi{max} 
$
and
$
 \Gamma^\rmi{\virt}_\rmi{max}
$. 
In some theories all reactions are inelastic, 
but the division is non-trivial if the system
displays a global or discrete symmetry, as is 
typically the case in dark matter models.  
It is sufficient if the symmetry is an approximate one,  
emerging for instance
in the non-relativistic limit. 

The inelastic and elastic processes proceed with very different
rates if $T \ll m^{ }_\aS$. 
In this situation 
$
  \Gamma^\rmi{phys}_\inel
$
is exponentially suppressed compared with 
$
  \Gamma^\rmi{phys}_\elas
$, 
{\it viz.} 
\be
 \Gamma^\rmi{phys}_\inel \sim e^{-m^{ }_\aS / T}\, \Gamma^\rmi{phys}_\elas
 \;, \quad
 T \ll m^{ }_\aS
 \;. \la{hierarchy} 
\ee
The reason is the appearance of an additional $\aS$ particle
in the initial or final state.

If the hierarchy in \eq\nr{hierarchy} is present, 
there is a temperature regime
in which the system should be in kinetic equilibrium, but out
of the chemical one. To describe such a system, we may take 
a momentum average, and adopt 
$n^{ }_\aS \equiv \int^{ }_\vec{k} f^{ }_\aS$
as the only non-equilibrium variable. 

To obtain an equation for $n^{ }_{\aS}$, 
we return to \eq\nr{boltzmann}, assume $f^{ }_\aS$ to be translationally
invariant, and divide by $\omega$. 
Working in an expanding background, 
the integral over $\vec{k}$
yields $(\partial^{ }_t + 3 H)\, n^{ }_\aS$
on the left-hand side.
A key point is that 
on the right-hand side, the division by $\omega$ and 
the integration over $\vec{k}$ imply that the matrix element squared is 
averaged over {\em all momenta}. 
Then we can symmetrize the average. In particular, 
for processes leading to the elastic part of \eq\nr{inel_elas}, 
we can exchange initial- and final-state momenta, symbolically as
\ba
 && 
 \int_{\vec{k}}
 \frac{1}{2\omega} 
 \int \! {\rm d}\Phi^{ }_{1 + m \to n} 
 \, 
 \bigl \{  
 f^{ }_{\aS}
 (1 + f^{ }_{b_{1\aS}})
 \times 
 f^{ }_{a_1} \cdots f^{ }_{a_m} 
 (1\pm f^{ }_{b_2}) \cdots (1\pm f^{ }_{b_n}) 
 \nn[1.5mm] & & \;  
 - \, 
 (1 + f^{ }_{\aS})
 f^{ }_{b_{1\aS}}
 \times 
 f^{ }_{b_2} \cdots f^{ }_{b_n} 
 (1\pm f^{ }_{a_1}) \cdots (1\pm f^{ }_{a_m})
 \bigr\}
 \nn[2mm]
 & 
 \underset{ a^{ }_{1}\cdots a^{ }_m \leftrightarrow 
            b^{ }_{2}\cdots b^{ }_n 
          }{  
  \overset{ \aS \leftrightarrow b^{ }_{1\aS} }{ = } } 
 & 
 \int_{\vec{k}}
 \frac{1}{2\omega} 
 \int \! {\rm d}\Phi^{ }_{1 + m \to n} 
 \, 
 \bigl \{  
 f^{ }_{b_{1\aS} }
 (1 + f^{ }_{\aS})
 \times 
 f^{ }_{b_2} \cdots f^{ }_{b_n} 
 (1\pm f^{ }_{a_1}) \cdots (1\pm f^{ }_{a_m}) 
 \nn[1.5mm] & & \;  
 - \, 
 (1 + f^{ }_{b_{1\aS}})
 f^{ }_{\aS}
 \times 
 f^{ }_{a_1} \cdots f^{ }_{a_m} 
 (1\pm f^{ }_{b_2}) \cdots (1\pm f^{ }_{b_n})
 \bigr\}
 \;. \hspace*{5mm} \la{step1}
\ea
The two terms 
are opposites of each other, and the result cancels by antisymmetry.
If the number of spectators changes, there are two different 
processes (spectators increase or decrease), each with their
own loss and gain terms. Then we may inspect the four processes
together, and the substitution $\aS\leftrightarrow b^{ }_{1\aS}$
alone shows that elastic processes drop out.\footnote{%
 As already mentioned,
 we assume 
 $
   c \sum_\rmii{spins} |\mathcal{M}|^2_{{\aS}+m\to n}
 $
 to be invariant in these exchanges. 
 } 

For the inelastic processes, the left-hand side of 
\eq\nr{step1} is replaced with 
(we show this with the example of two dark matter particles 
in the initial and none in the final state, 
however this can be generalized, see below)
\ba
 && 
 \int_{\vec{k}}
 \frac{1}{2\omega} 
 \int \! {\rm d}\Phi^{ }_{1 + m \to n} 
 \, 
 \bigl \{  
 f^{ }_{\aS}
 f^{ }_{a_{1\aS}}
 \times 
 f^{ }_{a_2} \cdots f^{ }_{a_m} 
 (1\pm f^{ }_{b_1}) \cdots (1\pm f^{ }_{b_n}) 
 \nn[1.5mm] & & \;  
 - \, 
 (1 + f^{ }_{\aS})
 (1 + f^{ }_{a_{1\aS}})
 \times 
 f^{ }_{b_1} \cdots f^{ }_{b_n} 
 (1\pm f^{ }_{a_2}) \cdots (1\pm f^{ }_{a_m})
 \bigr\}
 \;. \la{step3} 
\ea
Let us analyze this in the linear response regime, writing 
$f^{ }_\aS \to \bar{f}^{ }_\aS + \delta f^{ }_\aS$. Then 
\ba
 f^{ }_{\aS} f^{ }_{a_{1\aS}} & \to & 
 \underbrace{
 \bar{f}^{ }_{\aS} \bar{f}^{ }_{a_{1\aS}}
            }_{\rmii{cancels by detailed balance}}
 + 
 \underbrace{ 
 \delta f^{ }_{\aS} \bar{f}^{ }_{a_{1\aS}} 
 + 
 \bar{f}^{ }_{\aS} \delta f^{ }_{a_{1\aS}}
           }_{\aS \leftrightarrow a_{1\aS}
              \;\Rightarrow\;
              2 \, \delta f^{ }_{\aS} \bar{f}^{ }_{a_{1\aS}} 
             }
 \quad + \quad  \rmO(\delta^2) 
 \;, \la{step4} \\[2mm] 
 (1 + f^{ }_{\aS})(1+ f^{ }_{a_{1\aS}}) & \to & 
 \underbrace{
 (1+\bar{f}^{ }_{\aS})(1+ \bar{f}^{ }_{a_{1\aS}})
            }_{\rmii{cancels by detailed balance}}
 + 
 \underbrace{ 
 \delta f^{ }_{\aS} (1+ \bar{f}^{ }_{a_{1\aS}}) 
 + 
 (1+\bar{f}^{ }_{\aS}) \delta f^{ }_{a_{1\aS}}
           }_{\aS \leftrightarrow a_{1\aS}
              \;\Rightarrow\;
              2 \, \delta f^{ }_{\aS} (1 + \bar{f}^{ }_{a_{1\aS}} )
             }
 \quad + \quad  \rmO(\delta^2) 
 \;. \la{step5} \nn 
\ea
In total we get 
\ba
 (\partial^{ }_t + 3 H) n^{ }_\aS & \approx & 
 -\sum_{m,n}
 \int_\vec{k} \frac{2\, \delta f^{ }_{\aS}}{\omega} \, \frac{1}{2}   
 \int \! {\rm d}\Phi^{ }_{1 + m \to n} 
 \\ && \hspace*{-2.5cm}
 \times
 \bigl \{  
 \bar{f}^{ }_{a_{1\aS}} \cdots f^{ }_{a_m} 
 (1\pm f^{ }_{b_1}) \cdots (1\pm f^{ }_{b_n}) 
 - 
 f^{ }_{b_1} \cdots f^{ }_{b_n} 
 (1 + \bar{f}^{ }_{a_{1\aS}}) \cdots (1\pm f^{ }_{a_m})
 \bigr\}
  \, 
 c \sum_\rmi{spins} |\mathcal{M}|^2_{{\aS}+m\to n}
 \nn & = & 
 -2 
 \int_\vec{k} \frac{\delta f^{ }_{\aS}}{\omega}  
 \, \omega \,\Gamma^\rmi{\real}_\rmi{max,inel}(k)
 \;, \la{rate_chem}
\ea
where we recognized an inelastic part of
the maximal interaction rate, defined in \eq\nr{phasespace}. 

Given that  
$
 \Gamma^\rmi{\real}_\rmi{max,inel}
$
is a well-defined subpart of 
$
 \Gamma^\rmi{\real}_\rmi{max}
$, 
we can consider the equally well-defined virtual correction
$
 \Gamma^\rmi{\virt}_\rmi{max,inel}
$. 
At NLO, their sum yields 
\be
 \Gamma^\rmi{phys}_\rmi{inel}(k)
 \; \equiv \; 
 \Gamma^\rmi{\real}_\rmi{max,inel}(k)
 + 
 \Gamma^\rmi{\virt}_\rmi{max,inel}(k)
 \;, 
\ee
in analogy with \eq\nr{max}.

We may now generalize the consideration of
\eqs\nr{step3}--\nr{rate_chem}.  
For a set of $i$ dark matter particles appearing on one side only, 
we may undertake a symmetrization like in 
\eqs\nr{step4} and \nr{step5}.
Then we may represent 
$ \Gamma^\rmi{phys}_\inel(k) $ as
\be
 \Gamma^\rmi{phys}_\inel(k)
 = 
 \sum_{i=1}^{\infty}  
 \Gamma^\rmi{phys}_\rmi{inel($i$)}(k)
 \;.
\ee
Because of the symmetrization, the rate 
equation \nr{boltzmann3} obtains a symmetry factor $i$. 

Finally, returning to \eq\nr{rate_chem} and 
assuming kinetic equilibrium, 
the deviation of the phase-space density can be expressed as 
\be
 \delta f^{ }_\aS(k) = \nB^{ }(\omega) \,
 \frac{\delta n^{ }_\aS}{n^{ }_\rmi{eq}}
 \;, \quad
 n^{ }_\rmi{eq} \equiv \int_\vec{k} \nB^{ }(\omega)
 \;, \quad 
 \delta n^{ }_\aS \; \equiv \; n^{ }_\aS - n^{ }_\rmi{eq}
 \;. \la{ansatz}
\ee
The weighting by $\nB^{ }(\omega)$ and the division by 
$n^{ }_\rmi{eq}$ in \eq\nr{ansatz} prompt us to define
\be
 \bigl\langle\, \Gamma^\rmi{phys}_\rmi{inel($i$)} \,\bigr\rangle 
 \; \equiv \; 
 \frac{\int^{ }_\vec{k} \Gamma^\rmi{phys}_\rmi{inel($i$)}(k)\, \nB^{ }(\omega)}
      {\int^{ }_\vec{k} \nB^{ }(\omega)}
 \;. \la{ave}
\ee
Thereby the momentum-averaged equation, expanded to 
first order in $\delta n^{ }_\aS$, turns into
\be
 \bigl( \partial^{ }_t + 3 H \bigr) n^{ }_\aS 
 = -\sum_{i=1}^{\infty} i    
 \, \bigl\langle\, \Gamma^\rmi{phys}_\rmi{inel($i$)} \,\bigr\rangle 
 \,\delta n^{ }_\aS
 \; + \; \rmO( \delta n^2_\aS )
 \;. \la{boltzmann3}
\ee
In models with a discrete or continuous symmetry, 
the leading contribution originates 
from $i=2$, and making use of 
$
 n^{ }_\aS - n^{ }_\rmi{eq} \approx 
 ( n^2_\aS - n^2_\rmi{eq} ) / (2 n^{ }_\rmi{eq})
$ as is valid in the linear response regime on which our
derivation relied, 
we find the usual evolution equation, 
\be
 \bigl( \partial^{ }_t + 3 H \bigr) n^{ }_\aS 
 \approx - 
 \langle \sigma \vrel \rangle 
 \bigl( n^2_\aS - n^2_\rmi{eq} \bigr)
 \;, \quad
 \langle \sigma \vrel \rangle 
 \;\equiv\; 
  \frac{
   \langle \Gamma^\rmi{phys}_\rmi{inel($2$)} \rangle
      }{n^{ }_\rmi{eq}}   
 \; = \; 
  \frac{
   \int_\vec{k}  \Gamma^\rmi{phys}_\rmi{inel($2$)} \, \nB^{ } 
      }{n^{2}_\rmi{eq}}   
 \;. \la{lw}
\ee

To summarize this section, 
we have argued that \eq\nr{phasespace} contains a part, 
namely processes that can be characterized as inelastic,  
cf.\ \eq\nr{inel_elas}, whose momentum average permits to define
a chemical equilibration rate
and a corresponding Boltzmann 
equation, cf.\ \eq\nr{boltzmann3}. The strength of this formulation
is that we can promote the rate coefficients to include
virtual corrections, according to \eq\nr{max}, 
which ensures the absence of 
mass singularities~\cite{kln1,kln2}. 
In the linear response regime and  
assuming the presence of kinetic equilibrium, 
the formalism reduces to the usual form,  
cf.\ \eq\nr{lw}. The coefficient
$ \langle \sigma \vrel \rangle $ incorporates the 
influence of fast processes, the functional form of 
\eq\nr{lw} those
of the slow variables. 
Thereby \eq\nr{lw} can normally 
also be applied once freeze-out has taken place, 
i.e.\ $n^{ }_\aS \gg n^{ }_\rmi{eq}$~\cite{clas2}.

%
\section{Matrix elements squared}
\la{se:MstarM} 

%
\subsection{Overview}
\la{ss:overview}

We have argued in the previous section that the dynamical 
information entering the dark matter evolution equation, 
cf.\ \eq\nr{lw}, can be obtained from a general class of 
thermally averaged rates, cf.\ \eq\nr{phasespace}, by 
restricting to inelastic processes and adding virtual corrections. 
For the class of \eq\nr{phasespace}, 
an algorithm has been worked out in which all relevant channels, 
and the virtual corrections that cancel mass singularities, can 
be derived from minimal information, contained in the decay 
matrix elements defined in \eq\nr{Theta_def}~\cite{phasespace}. 
In this section, we illustrate how the procedure works, 
by introducing a specific model (cf.\ \se\ref{se:model}). 

As far as the notation goes, we employ thermal averages 
like in \eq\nr{scat}, specifically 
$\scat{1\to n}(b^{ }_1,...,b^{ }_n)$. Kinematic 
invariants are defined in the usual way, 
\be
 \s{ij}^{ } \; \equiv \;
 (\P^{ }_{b_i} + \P^{ }_{b_j})^2
 \;. 
\ee

For deriving matrix elements 
in which one $\aS$ is in the initial state and the rest of them
appear in the final state, it is convenient to shift $\aS \to \aS + \bS$ 
in the Lagrangian, and treat $\bS$ as
a thermalized final-state field.
This simplifies the computation of combinatorial factors, 
and makes explicit the linear response philosophy. 

%
\subsection{Model and parameters}
\la{se:model}

We illustrate the procedure of \se\ref{se:boltzmann} with
the scalar singlet extension of the Standard Model 
(cf.,\ e.g.,\ refs.~\cite{singlet1,singlet2,singlet3,singlet4}
and references therein), 
defined by 
\be
 \mathcal{L} \; = \; \mathcal{L}^{ }_\rmii{\it SM}
  + \, \biggl\{
     \frac{1}{2} \partial^{\mu}_{ }\varphi \, \partial^{ }_\mu\varphi
  - \, \biggl[
                 \frac{1}{2} \, m_{\aS 0}^2\, \varphi^2
               + \frac{1}{2} \, \kappa\,  \varphi^2 H^\dagger H
               + \frac{1}{4} \, \lambda^{ }_\aS\, \varphi^4
    \,\biggr]\,  \biggr\} 
 \;, \la{L}
\ee
where $H$ is the Higgs doublet and an {\em ad hoc}
$\mathbbm{Z}$(2) symmetry 
has been imposed in order to reduce the number of parameters. 
After electroweak symmetry breaking, 
the Higgs doublet is parametrized as  
\be
 H = \frac{1}{\sqrt{2}}
 \biggl( 
 \begin{array}{c} 
  \phi^{ }_2 + i \phi^{ }_1 \\ 
  v + h - i \phi^{ }_3 
 \end{array} 
 \biggr) 
 \;, 
\ee
where $v|^{ }_{T=0}\simeq 246$~GeV, $h$ denotes the physical Higgs field, 
and the Goldstone modes $\phi^{ }_a$ are 
numbered in analogy with the Pauli matrices.  
The scalar singlet is assumed to be 
in its unbroken phase throughout the cosmic history, 
i.e.\ $ m_{\aS 0}^2 > 0 $, 
and its tree-level vacuum mass is given by 
$m_\aS^2 = m_{\aS 0}^2 + \kappa v^2 / 2$.\footnote{%
 At high temperatures, the singlet mass squared
 experiences thermal corrections, 
 first of all due to 
 $v^2 |^{ }_\T$, but also due to 
 other effects, if $\pi T \gg \min\{ m^{ }_\aS, m^{ }_h \}$.
 However, since we assume coupling strengths such 
 that freeze-out takes place deep in the 
 non-relativistic regime, with 
 $\pi T \ll \min\{ m^{ }_\aS, m^{ }_h \}$, 
 these are unimportant for us 
 and omitted for simplicity. If the vacuum value 
 of $m^{ }_\varphi$ were precisely known, however, 
 even small effects could have an impact,
 due to the vicinity of the kinematic threshold at 
 $m^{ }_\varphi \simeq m^{ }_h/2$. 
 } 


On the Standard Model side, an important role 
is played by the charm and bottom quarks 
and the strong gauge coupling. The values of the charm and 
bottom masses are conventionally tabulated at a renormalization 
scale $\bmu \simeq 2$~GeV, and we evolve them 
to a thermal scale $\bmu \simeq 2\pi T$.
In addition, we scale the quark masses by the temperature dependence
of the Higgs expectation value, 
$
 v|^{ }_\T \simeq v|^{ }_0 \re\sqrt{1 - T^2/\Tc^2}
$, 
where the pseudocritical temperature $\Tc\approx 160$~GeV
can be adopted from ref.~\cite{dono}. Of course, at
the temperatures $T \sim (1-3)$~GeV that are most important
for us, the latter effect is minuscule. 

As far as the top quark is concerned, it can be integrated out 
deep in the Higgs phase, which yields the effective operator~\cite{dim5}
\be
 \mathcal{L} \; \supset \; 
 - \frac{g_3^2}{(4\pi)^2}
   \frac{h\, G^a_{\mu\nu} G^{a\mu\nu}_{ }}{3v}
 \;, \la{dim5}
\ee
where $g_3^2 \equiv 4\pi\alphas^{ }$ is the strong gauge coupling
and $G^a_{\mu\nu}$ is the SU(3) field strength tensor. 
We fix 
$\alphas^{ }(m^{ }_\aZ) \approx 0.118 $, 
and again evolve this to $\bmu \simeq 2\pi T$. 
The role that \eq\nr{dim5} plays for Higgs physics 
at temperatures of a few GeV has been elaborated upon 
in ref.~\cite{gw}. 

For reference, 
let us start by briefly considering $T > 160$~GeV, where 
electroweak symmetry is restored, 
{\it viz.} $v|^{ }_\T \simeq 0$. This regime may play a role
for freeze-in dark matter production, and also offers for 
a partial crosscheck of matrix elements squared, 
by their continuity. 
A relatively straightforward computation yields
\ba
  \omega\,
  \Gamma^\rmi{\real}_{1\to 3} & = & 
  2 \kappa^2\,
   \scat{1 \to 3}(\bS,\phi,\phi) \,
 \; + \; 
   6 \lambda^{2}_\aS\,  
   \scat{1 \to 3}(\bS,\bS,\bS) \,
 \;, \la{nlo_gauge_symmetric}
\ea
where the first term represents equilibration through Higgs scatterings,  
and the latter through $\aS$ self-interactions. By $\phi$ we have denoted
a Standard Model scalar particle (with 4 real degrees of freedom) 
in the absence of electroweak symmetry breaking. As stressed before, 
even if the $1\to 3$ decays in \eq\nr{nlo_gauge_symmetric} are 
kinematically forbidden, their algebraic forms capture 
the matrix elements squared of all allowed crossed channels. 
When we go to the
Higgs phase, \eq\nr{nlo_gauge_symmetric}
is replaced by an expression 
containing contributions from many channels, and we now turn
to which of them are the most important ones. 

%
\subsection{Which processes are important?}

%
\begin{figure}[t]
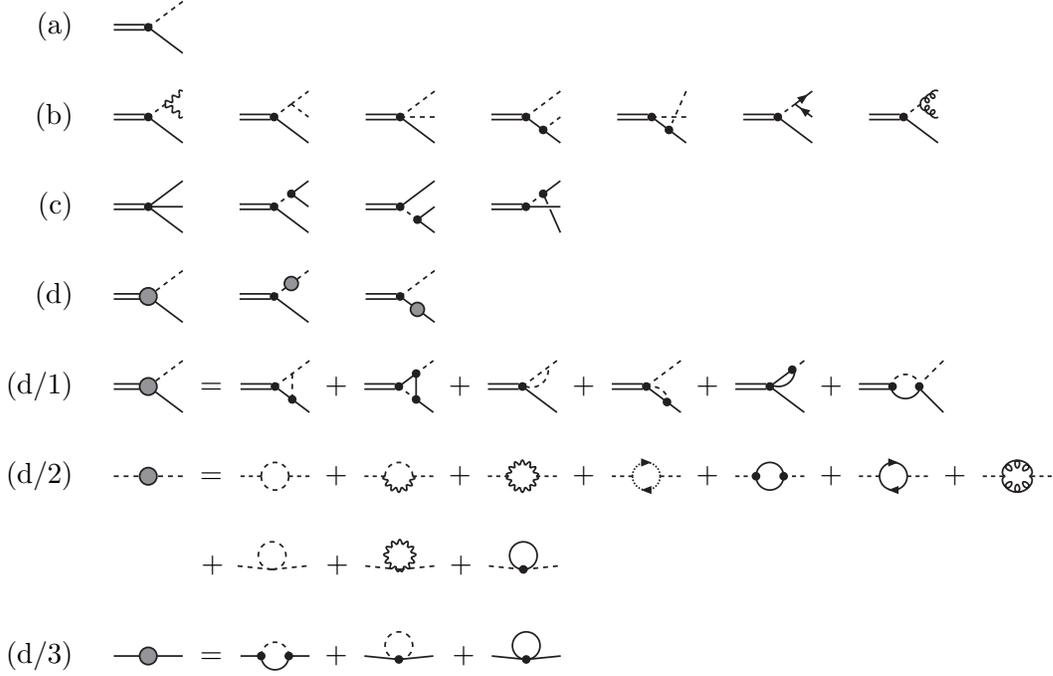


\vspace*{-5mm}

\begin{eqnarray*}
\mbox{(a)} && \hspace*{-0.8cm} 
 \Ampl  \hspace*{-1.0cm}
 \\ 
\mbox{(b)} && \hspace*{-0.8cm} 
 \AmplE  \hspace*{-1.0cm}
 \AmplG  \hspace*{-1.0cm}
 \AmplH  \hspace*{-1.0cm}
 \AmplI  \hspace*{-1.0cm}
 \AmplJ  \hspace*{-1.0cm}
 \AmplD  \hspace*{-1.0cm}
 \AmplF  \hspace*{-1.0cm}
 \\ 
\mbox{(c)} && \hspace*{-0.8cm} 
 \AmplK  \hspace*{-1.0cm}
 \AmplL  \hspace*{-1.0cm}
 \AmplM  \hspace*{-1.0cm}
 \AmplN  \hspace*{-1.0cm}
 \\ 
\mbox{(d)} && \hspace*{-0.8cm} 
 \AmplO  \hspace*{-1.0cm}
 \AmplP  \hspace*{-1.0cm}
 \AmplQ  \hspace*{-1.0cm}
 \\ 
\mbox{(d/1)} && \hspace*{-0.8cm} 
 \AmplO  \hspace*{-1.0cm} = \hspace*{-0.5cm}
 \VertexA  \hspace*{-1.0cm} + \hspace*{-0.5cm}
 \VertexB  \hspace*{-1.0cm} + \hspace*{-0.5cm} 
 \VertexD  \hspace*{-1.0cm} + \hspace*{-0.5cm} 
 \VertexC  \hspace*{-1.0cm} + \hspace*{-0.5cm} 
 \VertexE  \hspace*{-1.0cm} + \hspace*{-0.5cm} 
 \VertexF  \hspace*{-1.0cm} 
 \\ 
\mbox{(d/2)} && \hspace*{-0.8cm} 
 \selfEA  \hspace*{-1.0cm} = \hspace*{-0.5cm}
 \scalarA  \hspace*{-1.0cm} + \hspace*{-0.5cm} 
 \scalarB  \hspace*{-1.0cm} + \hspace*{-0.5cm} 
 \scalarC  \hspace*{-1.0cm} + \hspace*{-0.5cm} 
 \scalarG  \hspace*{-1.0cm} + \hspace*{-0.5cm} 
 \scalarD  \hspace*{-1.0cm} + \hspace*{-0.5cm} 
 \scalarE  \hspace*{-1.0cm} + \hspace*{-0.5cm} 
 \scalarF  
 \\ 
  && \hspace*{1.0cm} + \hspace*{-0.45cm} 
 \scalarH  \hspace*{-0.95cm} + \hspace*{-0.5cm} 
 \scalarI  \hspace*{-1.0cm} + \hspace*{-0.5cm} 
 \scalarJ  
 \\ 
\mbox{(d/3)} && \hspace*{-0.8cm} 
 \selfEB  \hspace*{-1.0cm} = \hspace*{-0.5cm}
 \singletA  \hspace*{-1.0cm} + \hspace*{-0.5cm} 
 \singletB  \hspace*{-0.95cm} + \hspace*{-0.5cm} 
 \singletC  
\end{eqnarray*}

\vspace*{-5mm}

\caption[a]{\small 
Amplitudes for the decay and/or 
production of a $\aS$-particle (denoted by a double line). 
Many of these processes are kinematically forbidden, but this is 
inessential, as we only use them for extracting matrix elements
squared; crossings lead to allowed processes, 
e.g.\ $2\leftrightarrow 2$ scatterings, or 
$1\leftrightarrow 2$ decays or inverse decays
between a Higgs boson and two $\aS$-particles. 
Dashed lines denote Higgs fields, wiggly lines
weak gauge bosons, arrowed lines fermions, curly lines gluons, and 
straight lines thermalized singlet modes, 
denoted by $\bS$ in \se\ref{se:MstarM}. 
Only physical particles (no Goldstones) are shown as final states. 
The small blobs denote non-Standard Model couplings,  
and the gray blobs virtual corrections.  
The amplitudes have been classified as: 
(a)~$1\to 2$ processes; 
(b)~$1\to 3$ processes with two singlets;
(c)~$1\to 3$ processes with four singlets; 
(d)~virtual corrections to $1\to 2$ processes.
} 
\la{fig:boltzmann}
\end{figure}
%

When $T < 160$~GeV, the Higgs mechanism is active.
Then \eq\nr{nlo_gauge_symmetric} splits into many individual
processes, and it is furthermore supplemented by
additional matrix elements squared, proportional to~$v$. 
The corresponding amplitudes are illustrated in 
\fig\ref{fig:boltzmann}. The expressions are
collected in appendix~A, and here we single out the crucial ones. 

We note, first of all, that in the temperature range of interest, 
$T\lsim 10$~GeV, many Standard Model particles are heavy
($m^{ }_{h,\;Z^0,\;W^\pm,\;t} \gg \pi T$).
{\em Real processes} containing such external states
are exponentially suppressed. 
At the same time, {\em virtual processes} involving these
particles are not small: they contain large logarithms. 
But they are small compared with the $1\leftrightarrow 2$ process, 
whose parameters they correct. 
As a leftover from correcting parameters, 
they also lead to higher-dimensional operators, the largest
of which was introduced in \eq\nr{dim5}. 

To summarize, 
the most important processes at $T\lsim 10$~GeV are the 
leading-order $1\leftrightarrow 2$ one; those $2\leftrightarrow 2$
processes which contain particles with a mass $\lsim \pi T$, like
charm and bottom quarks, 
as these are not exponentially suppressed; 
as well as $2\leftrightarrow 2$ processes
originating through the higher-dimensional operator
in \eq\nr{dim5}.

%
\section{Numerical results}
\la{se:numerics} 

Given the matrix elements squared 
from appendix~A, the algorithm
of ref.~\cite{phasespace} determines all crossed processes
(specifically the $2\to 2$ and $3\to 1$ ones) contributing
to the full rate in \eq\nr{phasespace}, as well as 
IR sensitive virtual corrections. 
According to 
\eqs\nr{inel_elas} and \nr{boltzmann3}, we subsequently  
select the inelastic reactions. Given that this subclass differs
parameterically from the elastic processes,
by $ \sim e^{-m^{ }_\aS / T} $, 
the cancellation of mass singularities remains guaranteed. 
After the inclusion of
the virtual corrections, 
we obtain the coefficients denoted by $\Gamma^\rmi{phys}_\inel$
in \eq\nr{inel_elas}. For simplicity we 
drop the superscript from the rate coefficients, 
employing from now on just $\Gamma^{ }_\inel$. 

On the side of technical details, we remark that second order
poles in matrix elements squared 
are treated as parametric derivatives of first order poles, 
and first order poles are regularized as principal values. 
The dependence on the regularization drops out when real and virtual
corrections are summed together. As for UV divergences
in the virtual corrections, they are related to the 
renormalization of the parameters appearing in the 
$1\leftrightarrow 2$ process. 
As already mentioned, 
a convenient choice 
is to set the renormalization scale to $\bmu \simeq 2\pi T$.
In practice, choices related to renormalization are numerically
insignificant in our example. 

In order to model confining effects influencing
charm and bottom quarks as well as gluons, 
we have adopted the phenomenological
replacement $\Nc^{ }\to N^{ }_{\rm c,eff} < 3$ from 
ref.~\cite{nuMSM}. However, we have also considered the 
non-interacting value $\Nc^{ } = 3 $, and indicate
the difference of the two prescriptions 
as an error band in \fig\ref{fig:Gammainel}. 

As for the parameters, the key choice is the value of the coupling 
$\kappa$ in \eq\nr{L}. We have varied it around the value leading to
the correct dark matter abundance, found to be  
\be
 \kappa^{ }_\rmi{dm[here]} \approx 0.00064
 \quad (m^{ }_\aS \approx 60~\mbox{GeV})
 \;. \la{kappa} 
\ee 
This is close to but slightly smaller than the values cited in 
refs.~\cite{tb,kk2} if kinetic equilibrium is assumed, 
$\kappa^{ }_{\rmi{dm}\rmii{\!\cite{tb}}} \simeq 0.00066$ and 
$\kappa^{ }_{\rmi{dm}\rmii{\!\cite{kk2}}} \simeq 0.00068$, 
respectively.\footnote{%
 The much larger couplings  
 also discussed in ref.~\cite{tb} correspond to their case `B' that
 omits the elastic channels involving charm and bottom quarks,
 so that the system is far from kinetic equilibrium. 
 }

\begin{figure}[t]

\hspace*{-0.1cm}
\centerline{%
 \epsfysize=7.5cm\epsfbox{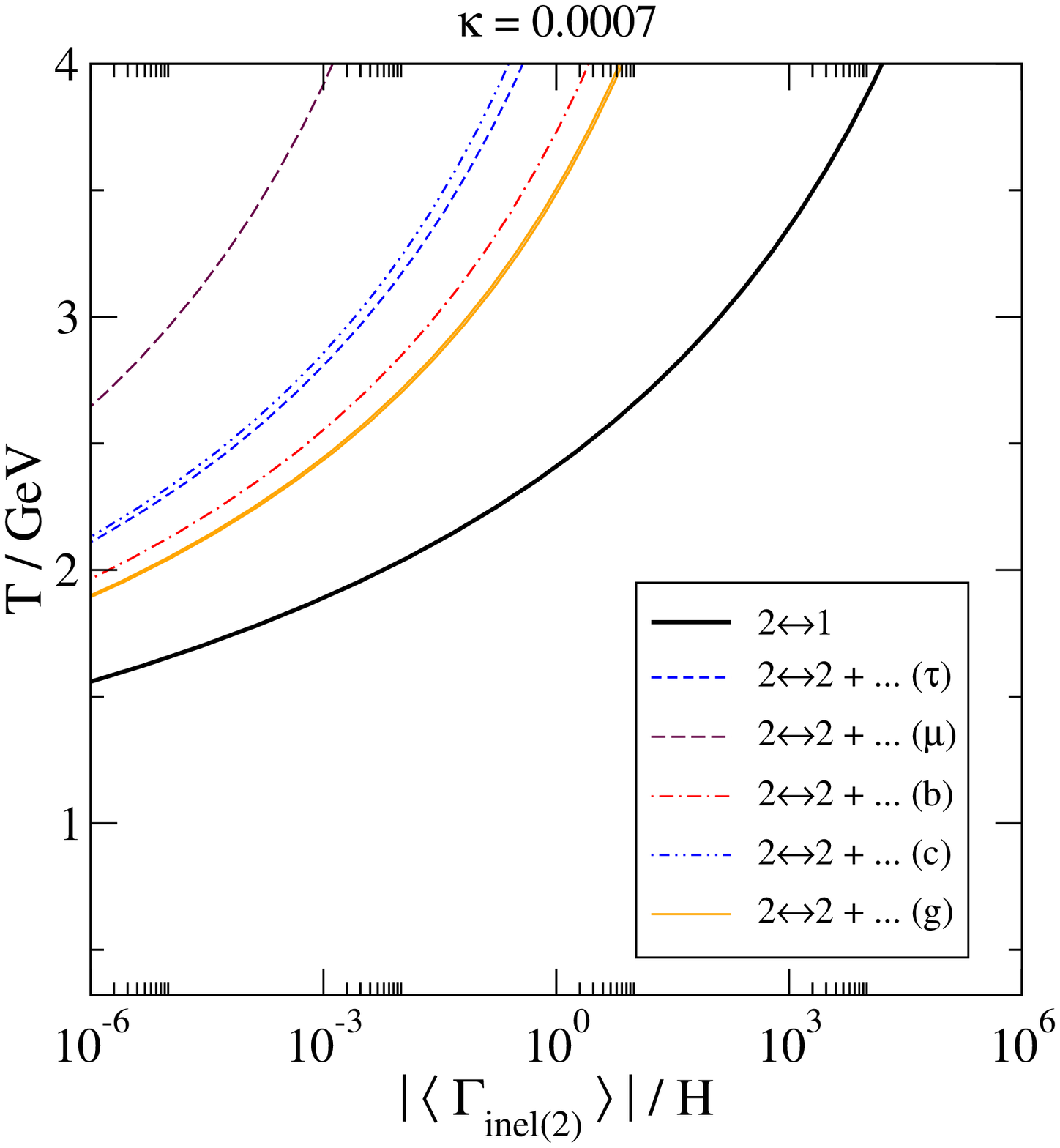}%
 \hspace{0.1cm}%
 \epsfysize=7.5cm\epsfbox{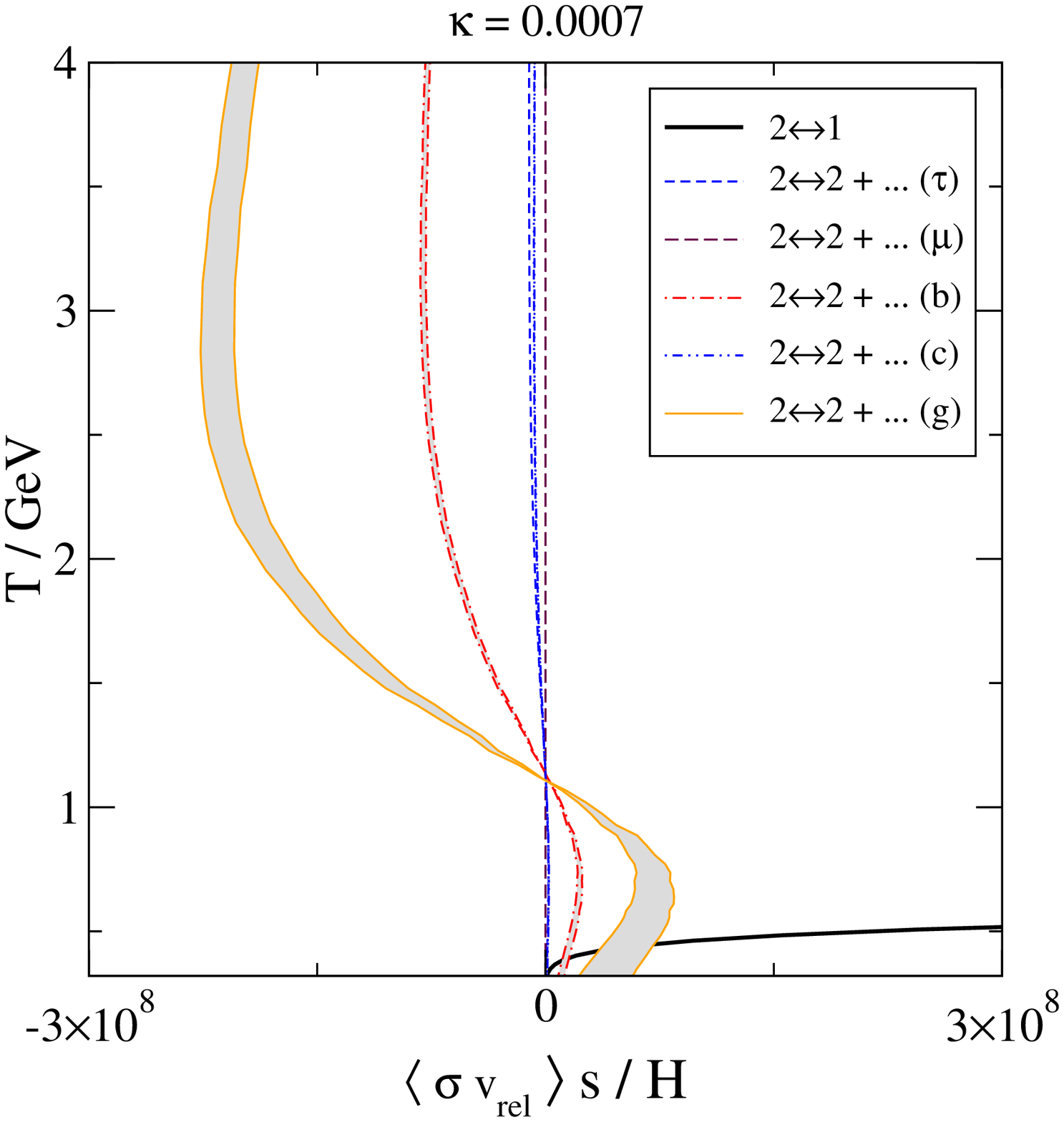}%
}

\caption[a]{\small
  Left: 
  the absolute value of the 
  momentum-averaged chemical equilibration rate from 
  \eq\nr{ave}, separated into contributions from different channels, 
  with ``$+...$'' indicating virtual corrections. 
  Freeze-out starts when 
  $
    \langle \Gamma^\rmi{ }_\rmi{inel($2$)} \rangle  \sim H
  $, 
  and in this regime NLO corrections are suppressed by
  orders of magnitude with respect to the  
  $2\leftrightarrow 1$ channel.  
  Right: 
  the coefficient $\langle \sigma \vrel \rangle$
  from \eq\nr{lw}, normalized as it appears in 
  \eq\nr{boltzmann5}
  (the factor $3 c_s^2 \simeq 1$ has been omitted here, though 
  it is included in our solution). 
  The purpose of this figure
  is to illustrate that NLO corrections can be negative, 
  because of their virtual part,
  and because the real part involves a principal value integral
  or its derivative, 
  however this only happens
  in a regime where the NLO corrections are utterly subdominant.
  The grey 
  bands indicate uncertainties in the evaluation of  
  hadronic contributions.   
}

\la{fig:Gammainel}
\end{figure}

In \fig\ref{fig:Gammainel}(left) different contributions to 
the absolute value of 
$
    \langle \Gamma^\rmi{ }_\rmi{inel($2$)} \rangle 
$
are plotted, 
normalized to the Hubble rate. The plot shows
that freeze-out must happen in the range $T \lsim 2$~GeV, 
when 
$
    \langle \Gamma^\rmi{ }_\rmi{inel($2$)} \rangle \lsim H
$,
and that NLO corrections are very small. To view the NLO 
corrections more clearly, we replot them in 
\fig\ref{fig:Gammainel}(right) in the combination  
$
 \langle \sigma \vrel \rangle s / H = 
 \langle \Gamma^\rmi{ }_\rmi{inel($2$)} \rangle s 
 / ( n^{ }_\rmi{eq} H)
$, 
where $s$ is the entropy density.

\begin{figure}[t]

\hspace*{-0.1cm}
\centerline{%
 \epsfysize=7.2cm\epsfbox{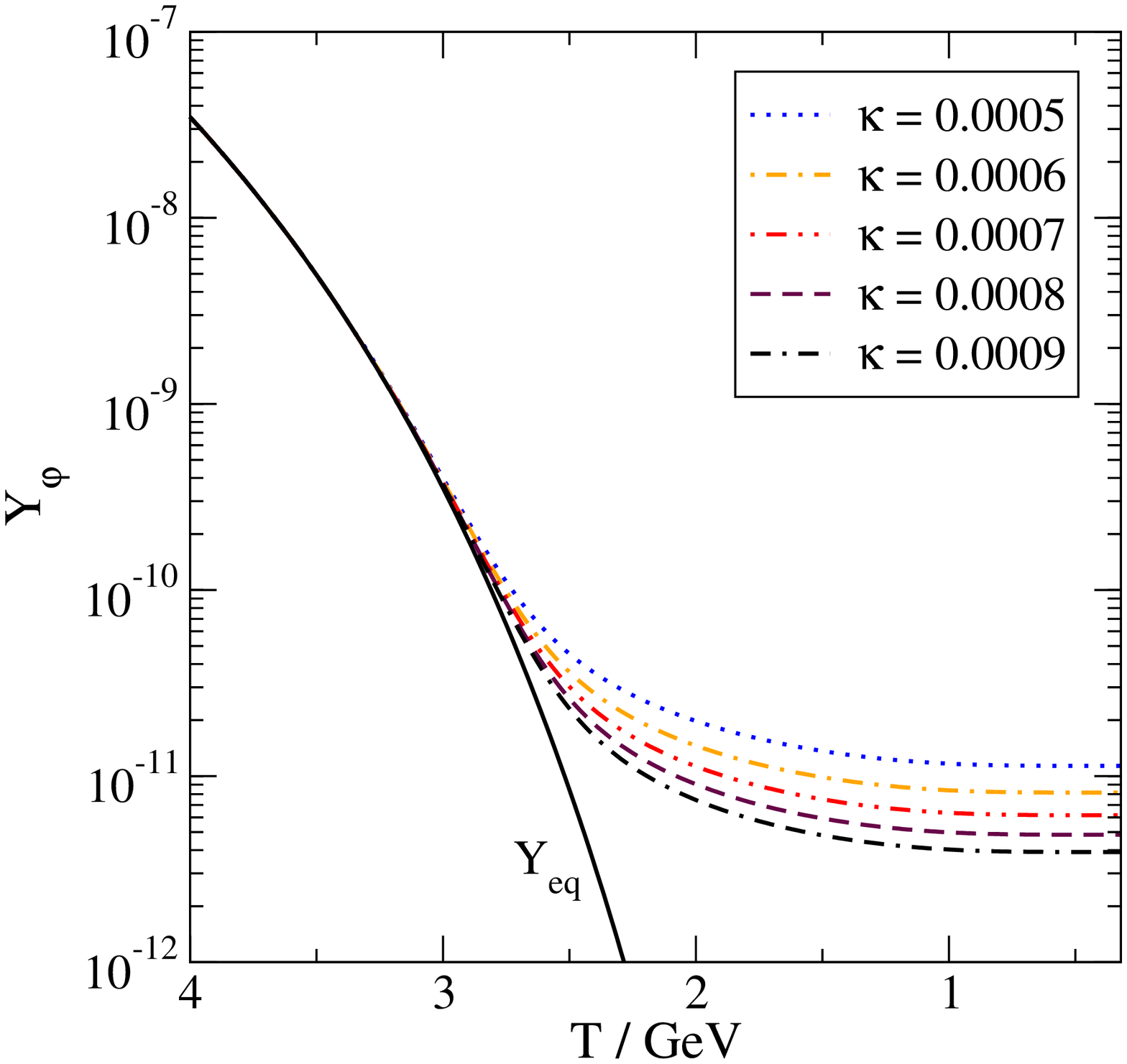}%
 \hspace{0.4cm}%
 \epsfysize=7.2cm\epsfbox{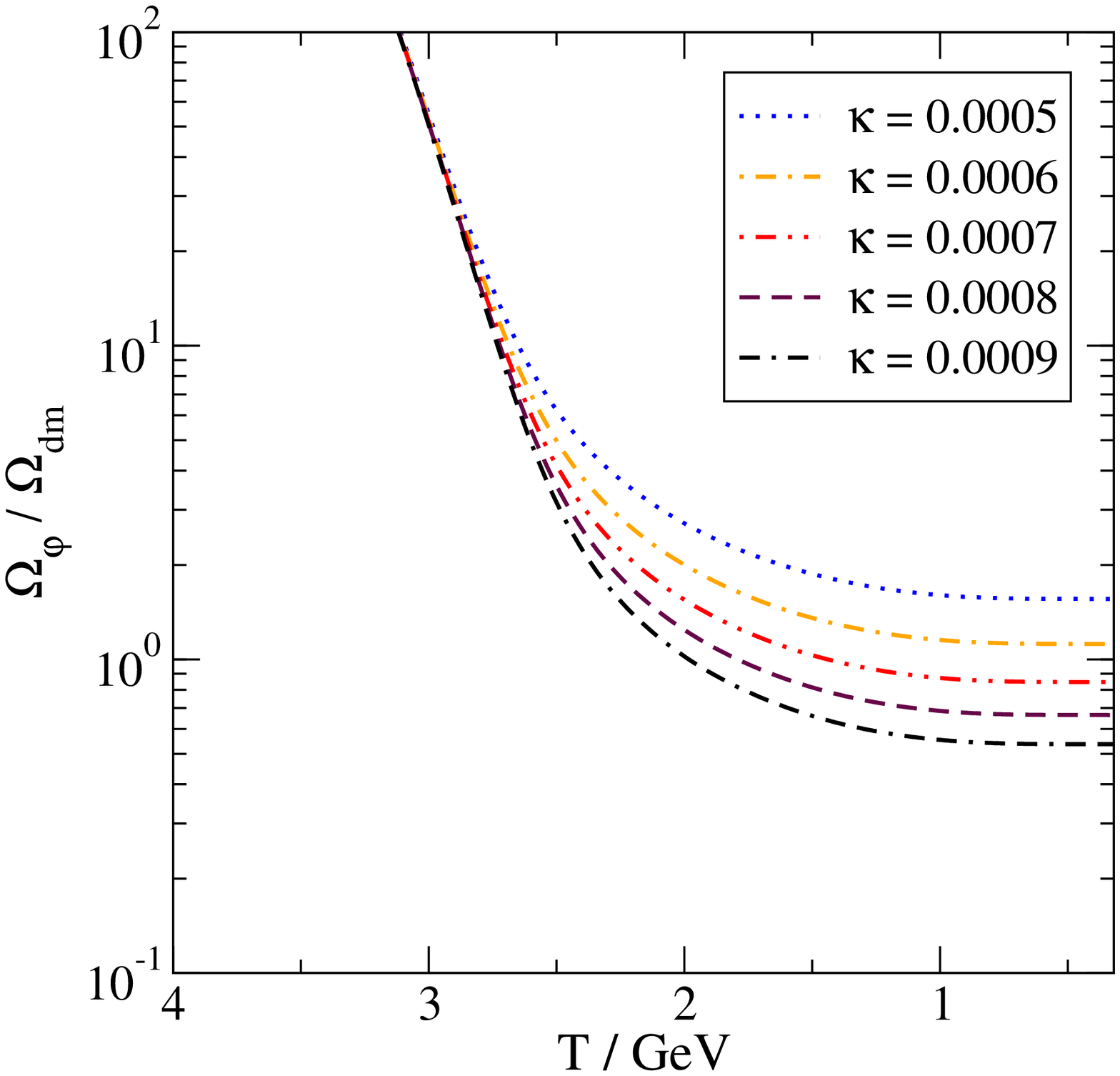}%
}

\caption[a]{\small
  Left: the solution of \eq\nr{boltzmann5} 
  for a few values of $\kappa$, 
  keeping the scalar singlet mass fixed at  
  $m^{ }_{\aS} \simeq 60$~GeV.
  Right: the same for the fractional contribution 
  of $\aS$ to the dark matter energy density.  
  The hadronic error bands from 
  \fig\ref{fig:Gammainel} are much narrower 
  than the line widths, given that 
  hadronic effects are much suppressed compared
  with the $2\leftrightarrow 1$ process
  when $T = (1-3)$~GeV.
  }

\la{fig:solution}
\end{figure}

Given the value of 
$
 \langle \sigma \vrel \rangle s  / H
$ 
from \fig\ref{fig:Gammainel}(right), 
we can integrate \eq\nr{lw}. 
In practice, it is convenient to normalize 
number densities by the entropy density $s$,  
denoting the corresponding yield parameters by
$Y^{ }_\aS \equiv n^{ }_\aS / s$, $Y^{ }_\rmi{eq} \equiv n^{ }_\rmi{eq}/s$. 
Introducing $x \equiv \ln (T^{ }_\rmi{max}/T)$ as an
integration variable,
and making use of the Jacobian
$
 {\rm d}x/{\rm d}t = 3 c_s^2 H
$, 
where $c_s^{2} = \partial p / \partial e$ is the speed of sound squared,  
\eq\nr{lw} turns into
\be
 \partial^{ }_x Y^{ }_\aS \approx
 - \frac{\langle \sigma \vrel \rangle s}{3 c_s^2 H}
 \, \bigl( Y_\aS^2 - Y_\rmi{eq}^2 \bigr)
 \;. \la{boltzmann5} 
\ee
For the thermodynamic functions 
$s$ and $c_s^2$ we adopt values derived
in refs.~\cite{ls,eos15}.\footnote{%
 These are tabulated at the web site 
 {\tt http://www.laine.itp.unibe.ch/eos15/}.
 }
{}From $Y^{ }_\aS$, the current energy density follows as 
$
 \Omega^{ }_\varphi / \Omega^{ }_\rmi{dm} 
 \approx 2.29 \, (m^{ }_\aS/\mbox{eV})\, Y^{ }_\varphi(x^{ }_\rmi{today})
$, 
where $\Omega^{ }_\rmi{dm} h^2 \approx 0.120$ refers to the 
observed value and $h$ is the reduced Hubble rate. 
The results obtained with 
the benchmark value $m^{ }_{\aS} \simeq$~60~GeV, 
both for $ Y^{ }_{\aS} $ and 
$
 \Omega^{ }_\varphi / \Omega^{ }_\rmi{dm} 
$, 
are shown in \fig\ref{fig:solution}.

%
\section{Summary and outlook}
\la{se:concl}

The common tool of dark matter computations, 
Boltzmann equations in their text-book form, 
do not incorporate thermal virtual corrections
(i.e.\ closed loops involving Bose and/or Fermi distributions), 
even though such effects could be important. 
On the formal level, they are necessary for  
cancelling mass singularities 
(mass thresholds or massless limits)
that originate from real scatterings. 
Physically, they lead to modified 
dispersion relations, and 
could then open up or 
close specific annihilation channels. 

The purpose of this paper has been to illustrate a way to
include thermal virtual corrections when dark matter annihilation is
influenced by an $s$-channel resonance. 
The treatment is simple-minded, both conceptually 
and technically. As a first step we 
carry out a systematic 
perturbative determination of the imaginary 
part of a retarded self-energy
in the linear response regime,\footnote{%
 In particular, resummations have not been touched upon, even though 
 they could become important if we go very close to the resonance
 production threshold. 
 } 
treating the 
narrow resonance as an on-shell particle at leading order. 
We found that a convenient way to implement this computation
is the substitution $ \aS \to \aS + \bS $, 
explained in \se\ref{ss:overview}. 
The finite width of the Higgs boson, modified by thermal corrections from 
Bose enhancement or Pauli blocking, originates as 
a part of NLO corrections, notably $2\leftrightarrow 2$ scatterings. 
But the finite width is not the only NLO correction: virtual 
corrections to the leading-order $2\leftrightarrow 1$ process
are of the same order and included at the same time.

If the physics that we are interested in takes place in 
the non-relativistic regime, and kinetic equilibrium can be assumed, 
the procedure can be completed by a second step. 
From the imaginary part of the retarded self-energy, 
we may identify the subpart    
that originates from inelastic reactions, and consider its 
momentum average. We have shown  that 
this reproduces the standard notion of 
the chemical equilibration rate (cf.\ \se\ref{se:chemical}), 
and ultimately leads to the usual
cosmological evolution equation (cf.\ \eq\nr{lw}). 
However, the overall philosophy 
should apply more generally than to systems in kinetic equilibrium, 
notably to freeze-in scenarios that operate in the relativistic
or ultrarelativistic regime
(cf., e.g., refs.~\cite{kin2,bg}), 
even if the practical implementation to such cases 
requires further consideration.

For an illustration, 
we returned to the well-studied example of 
scalar singlet dark matter,
assuming again that  
kinetic equilibrium is maintained by elastic scatterings. 
Then all NLO corrections are small
(cf.\ \fig\ref{fig:Gammainel}),  
and the thermally averaged cross section can be computed analytically 
(cf.\ \eq\nr{analytic}).
Numerically,
these results are in good agreement with 
previous literature (cf.\ \se\ref{se:numerics}), 
which relied on more complicated computations. 

Finally, in view of intensive 
discussions of the topic~\cite{tb,kk,drake,abe,kin,kk2}, 
we would like to put forward
one possibility for investigating kinetic non-equilibrium
in the non-relativistic regime. 
This is the use of Langevin simulations for determining 
momentum distributions. The Langevin 
description assumes that kinetic equilibrium is established by 
elastic scatterings and that the corresponding scattering rate 
is much smaller than the typical plasma interaction rates, i.e.\ 
that the dark matter particles are weakly coupled. But it does 
{\em not} assume that the plasma particles are weakly coupled
among themselves. Therefore it permits for the inclusion of NLO~\cite{sch}
or even non-perturbative information on the plasma
interactions~\cite{eucl}, 
as is certainly desirable for strongly interacting particles at 
$T \simeq (1-3)$~GeV. We note that such frameworks have been
widely applied for understanding the kinetic equilibration of
charm and bottom quarks in the heavy ion collision context~\cite{mt}. 

%
\section*{Acknowledgements}

I thank Simone Biondini, 
Torsten Bringmann and Kimmo Kainulainen 
for helpful discussions, 
Kalle Ala-Mattinen for providing numerical data from ref.~\cite{kk2}, 
and the University of Jyv\"askyl\"a for hospitality 
in September--November 2021, when this work got under way. 
My research was partly supported by the Swiss National Science Foundation
(SNSF) under grant 200020B-188712.

%
\appendix
\renewcommand{\thesection}{\Alph{section}} 
\renewcommand{\thesubsection}{\Alph{section}.\arabic{subsection}}
\renewcommand{\theequation}{\Alph{section}.\arabic{equation}}

%
\section{Details of matrix elements squared}

The purpose of this appendix is to list the matrix elements squared
originating from the reactions shown in \fig\ref{fig:boltzmann}. 
We recall that the optimal procedure is to determine
the algebraic structures of 
would-be $1\to 2$ and $1\to 3$ rates,
even if these are kinematically forbidden in practice. 
The other real processes ($2\to 1$, $2\to 2$, $3\to 1$) 
can then be generated by crossings, 
whereas the IR-sensitive virtual corrections
to $1\leftrightarrow 2$ are obtained
by finding the poles and residues appearing
in the matrix elements squared~\cite{phasespace}. 

%
\subsection{Gauge and scalar effects}

The leading-order diagram, shown in \fig\ref{fig:boltzmann}(a), yields
\be
  \omega\,
  \Gamma^\rmi{\real}_{1\to 2} \; = \; 
  \kappa^2 v^2 \,
   \scat{1 \to 2}(\bS,h) \,
  \;, \la{lo}
\ee
where
$
 \scat{1 \to 2}
$
corresponds to the notation introduced in \eq\nr{scat}.

The next-to-leading order contributions from 
\fig\ref{fig:boltzmann}(b) are suppressed by $\sim g^2$
with respect to \eq\nr{lo}, where $g^2$ is a generic Standard Model coupling. 
As a crosscheck, we have computed them in two different ways. 
One goes through the usual 
$
 \sum_\rmi{spins} |\mathcal{M}|^2_{ } 
$, 
cf.\ \eq\nr{Theta_def}, 
keeping only physical states as external particles and recalling
that the polarization sum for a massive gauge boson of mass $m$
and four-momentum $\P$ reads 
$
 \sum_{\lambda}^{ } 
 \epsilon^{\mu}_{\lambda} \epsilon^{\nu *}_{\lambda}
 = 
 -\eta^{\mu\nu}_{ } + \P^{\mu}_{ }\P^{\nu}_{ }/ m^2 
$.
The other method proceeds 
by computing the 2-loop self-energy of the decaying
particle, and extracting its cut. A benefit of the latter approach
is that it can straightforwardly
be carried out in a general $R^{ }_\xi$ gauge, 
including ghosts, and that subsequently the gauge independence 
of the result can be verified. 
Both methods yield the same results. 

Re-expressing subsequently $g^2 v^2$ as a mass squared,
the final expression of $\rmO(\kappa^2)$ can be put in the form
\ba
  \omega\,
  \Gamma^\rmi{\real}_{1\to 3} & \supset &
  \kappa^2\,\biggl\{\, 
   \scat{1 \to 3}(\bS,h,h) \,
   \frac{(\s{23}^{ } + 2 m_h^2)^2}{2(\s{23}^{ } - m_h^2)^2}  
 \nn[1mm] 
 & & 
  \quad\; + \,  
   \scat{1 \to 3}(\bS,Z^0,Z^0) \,
   \frac{(\s{23}^{ } - 2 m_\aZ^2)^2 + 8 m_\aZ^4}{2(\s{23}^{ } - m_h^2)^2}  
 \nn[1mm] 
 & & 
  \quad\; + \,  
   \scat{1 \to 3}(\bS,W^+,W^-) \,
   \frac{(\s{23}^{ } - 2 m_\aW^2)^2 + 8 m_\aW^4}{(\s{23}^{ } - m_h^2)^2}  
  \biggr\}
 \;. \la{nlo_gauge_broken} 
\ea
Going towards the symmetric phase, so that the masses go to zero, 
this agrees with the first part of \eq\nr{nlo_gauge_symmetric}.

As shown by the diagrams in \figs\ref{fig:boltzmann}(b,c), 
there are 
also amplitudes that are quadratic in the coupling $\kappa$
($\kappa$ is indicated by the small blob). 
For the matrix element
squared, this produces interference terms that are cubic in $\kappa$, 
\be
  \omega\,
  \Gamma^\rmi{\real}_{1\to 3} \; \supset \; 
  2 \kappa^3 v^2 \,
   \scat{1 \to 3}(\bS,h,h) \,
   \frac{\s{23}^{ } + 2 m_h^2}
        { (\s{12}^{ } - m_\aS^2) (\s{23}^{ } - m_h^2) }
 \;, \la{nlo_cubic}
\ee
as well as quartic dependences, 
\ba
  \omega\,
  \Gamma^\rmi{\real}_{1\to 3} & \supset & 
  \kappa^4 v^4 \, \biggl\{ \, 
   \scat{1 \to 3}(\bS,h,h) \,
   \biggl[
     \frac{1}{(\s{12}^{ } - m_\aS^2)^2} 
       + 
     \frac{1}{(\s{12}^{ } - m_\aS^2) 
              (\s{13}^{ } - m_\aS^2)} 
   \biggr] 
 \la{nlo_quartic} \\ 
 & & 
  \qquad\; + \,   
   \scat{1 \to 3}(\bS,\bS,\bS) \,
   \biggl[ 
     \frac{1}{2(\s{12}^{ } - m_h^2)^2} 
       + 
     \frac{1}{(\s{12}^{ } - m_h^2)
              (\s{13}^{ } - m_h^2)} 
   \biggr] 
   \biggr\} 
 \;. \vspace*{6mm} \nonumber 
\ea
Likewise, there are terms
with one or two appearances of $\lambda^{ }_\aS$, 
\ba
  \omega\,
  \Gamma^\rmi{\real}_{1\to 3} & \supset & 
   6 \lambda^{ }_\aS\, 
   \biggl\{ \, 
    \frac{ \kappa^2 v^2 }  
         { \s{12}^{ } - m_h^2 } 
  + 
   \lambda^{ }_\aS
   \, \biggr\} \,   
   \scat{1 \to 3}(\bS,\bS,\bS) \,
 \;, \la{nlo_lam2_broken}
\ea
but they only contribute to kinetic equilibration. 
The last term differs from 
\eq\nr{nlo_gauge_symmetric} in that the 
mass of the $\varphi$-particle has changed 
through the Higgs mechanism.

%
\subsection{Leptonic and hadronic effects}

At low temperatures, $\pi T \ll \min\{ m_\aW^{ }, m_h^{ }, m^{ }_\aS \}$, 
the most important effects originate from fermionic channels. 
As long as we are in the deconfined phase ($T \gg 160$~MeV), the 
contributions of leptons and light quarks 
(at leading order in $\alphas^{ }$)
amount to
\ba
  \omega\,
 \Gamma^\rmi{\real}_{1\to 3} & \supset &
  \kappa^2\, 
  \sum_{\ell}
   \scat{1 \to 3}(\bS,\ell,\bar\ell) \,
   \frac{2 m_\ell^2 (\s{23}^{ } - 4 m_\ell^2)}{(\s{23}^{ } - m_h^2)^2}  
 \nn 
 & + & 
  \kappa^2 \Nc^{ }\, 
  \sum_{q}
   \scat{1 \to 3}(\bS,q,\bar q) \,
   \frac{2 m_q^2 (\s{23}^{ } - 4 m_q^2)}{(\s{23}^{ } - m_h^2)^2}  
 \;, 
\ea
where $\ell\in\{e,\mu,\tau\}$ 
enumerates the charged leptons, and $q\in\{u,d,s,c,b,t\}$ 
the quarks
with $m^{ }_q \ll \max\{ \pi T, \mS^{ }/2 \}$. Quarks heavier
than this can be integrated out, 
yielding the higher-dimensional operator
in \eq\nr{dim5}. 
This gives the loop-suppressed contribution 
\be
  \omega\,
 \Gamma^\rmi{\real}_{1\to 3} \; \supset \;
 \frac{4 \kappa^2 a_s^2 (\Nc^2 - 1) }{9}
  \, \scat{1 \to 3}(\bS,g,g) \, 
   \frac{\s{23}^2}{(\s{23}^{ } - m_h^2)^2}   
 \;,  
\ee
where 
$
 a^{ }_s \equiv \alphas^{ }/\pi \equiv g_3^2 / (4\pi^2)
$
and $g$ stands for a gluon.
If we instead go to $T \ll 160$~MeV, hadronic degrees of freedom are
represented by pions. Treating them as degenerate for simplicity, 
we find
\be
 \omega\,
 \Gamma^\rmi{\real}_{1\to 3} \; \supset \;
  \frac{ \kappa^2 (\Nf^2 - 1) }{2} \, 
   \scat{1 \to 3}(\bS,\pi,\bar\pi) \, 
   \frac{m_\pi^4}{(\s{23}^{ } - m_h^2)^2}   
 \;, 
\ee
where $\Nf^{ } = 2$ is the number of light flavours. 

%
\section{Phase space integrals for the leading-order process}

We show in this appendix how the phase space integrals 
corresponding to the leading-order process, 
cf.\ \eq\nr{lo}, can be carried out analytically
in the non-relativistic limit $\pi T \ll m^{ }_{\aS}$. 

Incorporating the crossed channels and denoting the corresponding
phase-space average by 
$ \scat{1 \leftrightarrow 2}(\bS,h) $~\cite{phasespace}, 
the physical 
$\aS\aS\to h$ 
annihilation and 
$h\to \aS\aS$
pair creation rates can be represented as
\be
  \omega\,
  \Gamma^\rmi{\real}_{1\leftrightarrow 2} \; = \; 
  \scat{1 \leftrightarrow 2}(\bS,h) \, \Theta^{ }_{1\leftrightarrow 2}
  \;, \quad
  \Theta^{ }_{1\leftrightarrow 2} \; \equiv \;   \kappa^2 v^2 
  \;. \la{lo_full}
\ee
We perform the phase-space integrals in the plasma rest
frame, with $(\omega \equiv \sqrt{k^2 + m_\aS^2},\vec{k})$ denoting
the four-momentum of one of the singlet scalars. 
For a momentum-independent $\Theta^{ }_{1\leftrightarrow 2}$ 
and denoting $\beta \equiv 1/T$, this yields~\cite{lpm} 
\ba
  \scat{1 \leftrightarrow 2}(\bS,h) & \stackrel{m^{ }_\aS <\, m^{ }_h/2 }{=} & 
  \frac{T}{16\pi k} 
  \ln
  \biggl\{\,
    \frac{1 - e^{-\beta\epsilon_h^- }}{ 1 - e^{-\beta \epsilon_h^+}}   
    \frac{1 - e^{-\beta (\epsilon_h^+ - \omega )}}
         { 1 - e^{-\beta (\epsilon_h^- - \omega )}}   
  \,\biggr\}
  ^{ }
  _{
     \epsilon_h^\pm =  \frac{m_h^2}{2m_\aS^2}
     \Bigl(\,
       \omega \,\pm\, k \sqrt{1-\frac{4m_\aS^2}{m_h^2}} 
     \,\Bigr)
   } \hspace*{5mm} 
  \nn 
  & \stackrel{\pi T \,\ll\, m^{ }_\aS }{\approx} & 
  \frac{T}{16\pi k} 
  \bigl\{\,
    e^{-\beta (\epsilon_h^- - \omega ) }
 -  e^{-\beta (\epsilon_h^+ - \omega ) }
  \,\bigr\}
  \;, \la{int_0}
\ea
where $\epsilon_h^\pm$ denote the maximal and minimal energies
of the Higgs boson (originating from different angular configurations)
when the momentum of one of the $\aS$ particles
has been fixed, 
and $\epsilon_h^\pm - \omega$ are those of the co-annihilation
partner. Expanding the exponentials in \eq\nr{int_0}
in $k / \omega < 1$ and employing $\omega$ as the integration variable,  
so that $k = \sqrt{\omega^2 - m_\aS^2}$,
momentum averaging yields
\ba
 \bigl\langle\, 
   \Gamma^\rmi{\real}_{1\leftrightarrow 2}
 \,\bigr\rangle 
 & \stackrel{\pi T \, \ll \, m^{ }_\aS}{\approx} & 
 \frac{1}{n^{ }_\rmi{eq}}
 \int_0^\infty \! \frac{{\rm d}k\, k^2}{2\pi^2} 
 \, 
 \frac{ e^{- \beta\omega }  
    }{\omega}
 \, 
 \scat{1 \leftrightarrow 2}(\bS,h)\, \Theta^{ }_{1\leftrightarrow 2}  
 \nn[2mm] 
 & 
 \underset{ \rmii{\nr{int_rep_1}} }{  
  \overset{ \rmii{\nr{int_0}} }{ \approx } } 
 & 
 \frac{m^{ }_\aS \kappa^2 v^2  T}{16\pi^3 n^{ }_\rmi{eq}}
 \sqrt{1-\frac{4m_\aS^2}{m_h^2}}
 \, 
 \sum_{n=0}^\infty
 \frac{1}{n!}
 \biggl( \frac{m_h^2 - 4 m_\aS^2}{4 m^{ }_\aS T} \biggr)^n_{ }
 K^{ }_{1+n}\biggl( \frac{m_h^2}{2 m^{ }_\aS T} \biggr)
 \hspace*{5mm} 
 \nn[2mm] 
 & \stackrel{\rmii{\nr{int_rep_2}}}{=} & 
 \frac{\kappa^2 v^2  T}{32\pi^3 n^{ }_\rmi{eq}}
 \sqrt{ {m_h^2} - {4m_\aS^2} }
 \, 
 K^{ }_{1}\biggl( \frac{m_h}{T} \biggr)
 \;, \hspace*{5mm} \la{int_1}
\ea
where $K^{ }_{\nu}$ is a modified Bessel function, and 
we made use of the integral representations
\ba
 K^{ }_\nu(z) & = & 
 \frac{\Gamma(\tfr{1}{2})}{\Gamma(\nu + \tfr{1}{2})}
 \biggl( \frac{z}{2} \biggr)^\nu_{ }
 \int_1^\infty \! {\rm d}t \, e^{-z t} (t^2_{ } - 1)^{\nu - \tfr{1}{2}}_{ }
 \la{int_rep_1}
 \\[2mm] 
 & = &  
 \frac{1}{2} 
 \biggl( \frac{z}{2} \biggr)^\nu_{ }
 \int_0^\infty \! \frac{{\rm d}t}{t^{\nu + 1}_{ }}
 \, 
 \exp\biggl( - t - \frac{z^2}{4t} \biggr)
 \;. \la{int_rep_2}
\ea
The equilibrium density can be written as
\be
 n^{ }_\rmi{eq} 
 \; = \; 
 \int_0^\infty \! \frac{{\rm d}k\, k^2}{2\pi^2} 
 \, \nB^{ }(\omega) 
 \; 
  \underset{ \rmii{\nr{int_rep_1}} }{  
  \overset{ \pi T \, \ll \, m^{ }_\aS }{ \approx } } 
 \; 
 \frac{m_\aS^2 T}{2\pi^2} K^{ }_2 
 \biggl( \, \frac{m^{ }_\aS}{T} \,\biggr)
 \;, 
\ee
where we again took $\omega$ as an integration variable.
The momentum-averaged cross section from \eq\nr{lw} becomes
\be
  \langle \sigma \vrel \rangle^{ }_{1\leftrightarrow 2} 
  \quad
  \equiv
  \quad
  \frac{
   \langle\, \Gamma^\rmi{\real}_{1\leftrightarrow 2} \,\rangle
      }{n^{ }_\rmi{eq}}   
 \quad
 \underset{ { \pi T \, \ll \, m^{ }_\aS  } }{  
  \overset{ { m^{ }_\aS \, < \, {m^{ }_h} / {2}  } }{ \approx } } 
 \quad
 \frac{\pi \kappa^2 v^2 
  \sqrt{{m_h^2} - {4m_\aS^2}}
 \, 
 K^{ }_{1}\bigl( \frac{m_h}{T} \bigr)
 }
  {8 m_\aS^4 T K^{2}_2 
  \bigl( \, \frac{m^{ }_\aS}{T} \,\bigr) }
 \;.  \la{analytic} 
\ee

\small{
%

}

\end{document}